\documentclass[conference]{IEEEtran}

\usepackage{cite}
\usepackage{amsmath,amssymb,amsfonts}
\usepackage{algorithm}
\usepackage{algorithmic}
\usepackage{graphicx}
\usepackage{textcomp}
\usepackage{xcolor}
\usepackage[utf8]{inputenc}
\usepackage{subcaption} 
\usepackage{booktabs}  
\usepackage{multirow}
\usepackage{siunitx}
\graphicspath{ {./} } 
\usepackage[hyphens]{url}
\usepackage{tabularx}
\usepackage{comment}
\usepackage{listings}

\def\BibTeX{{\rm B\kern-.05em{\sc i\kern-.025em b}\kern-.08em
T\kern-.1667em\lower.7ex\hbox{E}\kern-.125emX}}

\begin{document}

\title{DriftSched: Adaptive QoS-Aware Scheduling under Runtime Token Drift for Multi-Tenant GPU Inference}

\author{\IEEEauthorblockN{Kathiravan Palaniappan}
\IEEEauthorblockA{\textit{Independent Researcher (University of Colorado Colorado Springs Alumni)} \\
kpalania@uccs.edu}
}

\maketitle

\begin{abstract}
Large Language Model (LLM) inference services increasingly operate on shared GPU infrastructure where heterogeneous requests compete for limited resources while requiring predictable latency and Quality-of-Service (QoS) guarantees. Although modern inference runtimes improve throughput through continuous batching and optimized memory management, accurately estimating workload cost at admission time remains challenging, and inaccurate estimates can lead to workload misclassification, queue imbalance, and increased tail latency.

This paper presents DriftSched, a workload-aware QoS scheduling framework for multi-tenant LLM inference serving. DriftSched combines admission-time workload characterization, tenant-aware queue management, token-budget estimation, and an online Exponential Moving Average (EMA)-based calibration mechanism that incorporates runtime execution feedback into future workload estimates. To study the impact of workload-estimation fidelity, we compare a lightweight whitespace-based proxy (\texttt{split()}) with tokenizer-aware accounting using the model's native tokenizer. Requests are organized into Premium, Standard, and Batch service tiers and evaluated under FIFO, Priority, Weighted, Shortest-Job-First (SJF), and Aging Priority scheduling policies.

Experimental evaluation under sustained GPU contention shows that workload-estimation fidelity significantly influences scheduling effectiveness. The EMA-based calibration mechanism compensates for systematic estimation errors introduced by whitespace-based workload characterization, enabling lightweight estimators to approach tokenizer-aware behavior after convergence. In contrast, tokenizer-aware accounting produces calibration factors that remain close to unity, indicating that most observed estimation drift originates from admission-time characterization error rather than intrinsic variability in model generation behavior. Across all evaluated configurations, scheduling policy selection has a greater impact on overall performance than runtime calibration alone. SJF achieves the strongest latency performance, reducing median end-to-end latency by approximately 42\% and P99 latency by approximately 16\% relative to FIFO, while Priority Scheduling provides the strongest tenant-level QoS differentiation. These findings suggest that accurate admission-time workload characterization is a key enabler of effective QoS-aware scheduling for multi-tenant GPU inference systems.
\end{abstract}


\section{Introduction}

The rapid adoption of Large Language Models (LLMs) has increased the demand for efficient multi-tenant GPU inference serving in modern AI datacenters. Enterprise AI platforms, cloud inference providers, and edge AI systems increasingly operate shared inference infrastructure where heterogeneous requests compete for limited GPU resources while requiring predictable latency, fairness, and Quality-of-Service (QoS) guarantees. Although modern inference runtimes such as vLLM~\cite{vllm2023}, TensorRT-LLM~\cite{tensorrtllm}, Orca~\cite{orca2022}, Sarathi~\cite{sarathi2023}, and SGLang~\cite{sglang} improve throughput through continuous batching and optimized memory management, efficient scheduling remains a fundamental challenge under sustained contention~\cite{transformer}.

Prior studies have examined deep-learning inference performance on modern CPU and GPU platforms~\cite{gdevai,deep-gap,nvidiat4,nvidial4}. While these works characterize architectural performance and scalability, they do not address how heterogeneous multi-tenant inference requests should be scheduled under contention. In practical LLM serving environments, overall performance depends not only on GPU capability but also on scheduling policy, workload-estimation accuracy, queue management, tenant prioritization, and batching dynamics.

A central requirement of any scheduling policy is the ability to estimate workload size before execution. Scheduling disciplines such as Shortest-Job-First (SJF), weighted scheduling, admission control, and priority-based queue management all rely on admission-time workload estimates to make informed placement decisions. In LLM serving environments, however, accurately characterizing workload size is challenging because runtime execution cost depends on prompt characteristics, generated output length, batching behavior, model-specific tokenization, and execution dynamics.

Many practical systems approximate workload cost using static token budgets, user-provided limits, or lightweight heuristics. Such approximations may introduce workload-characterization errors that propagate directly into scheduling decisions. Requests incorrectly classified as short may delay latency-sensitive workloads, while overestimating execution cost may unnecessarily defer lightweight requests. Under multi-tenant GPU contention, these admission-time estimation errors can accumulate into queue imbalance, fairness degradation, increased tail latency~\cite{tailatscale}, and degraded QoS.

This observation motivates a broader question:

\emph{How sensitive are modern QoS scheduling policies to workload-characterization fidelity?}

To investigate this question, we evaluate two admission-time workload-characterization strategies. The first utilizes a lightweight whitespace-delimited proxy (\texttt{split()}) that estimates workload size using word counts rather than model-native tokenization. While computationally inexpensive, such approximations may introduce systematic workload-estimation inaccuracies. The second employs tokenizer-aware accounting using the model's native tokenizer, providing a more accurate representation of execution cost in token space. Comparing these configurations enables direct evaluation of how workload-estimation fidelity influences scheduling behavior, queue dynamics, fairness, and latency under realistic multi-tenant inference workloads.

To support this study, we present \emph{DriftSched}, a workload-aware QoS scheduling framework for multi-tenant LLM inference serving. DriftSched combines admission-time workload characterization, tenant-aware queue management, workload estimation, and an optional Exponential Moving Average (EMA)-based feedback mechanism for adaptive calibration. Requests are organized into Premium, Standard, and Batch service tiers and evaluated under FIFO, Priority Scheduling, Weighted Scheduling, Shortest-Job-First (SJF), and Aging Priority Scheduling.

DriftSched optionally incorporates runtime feedback to adapt workload estimates and evaluate scheduling behavior under both approximate and tokenizer-aware characterization.

Experimental results show that workload-characterization fidelity affects scheduler behavior, although scheduling-policy selection has a larger impact on latency and QoS. SJF achieves the lowest latency, while Priority Scheduling provides the strongest tenant differentiation.

This paper makes the following contributions:

\begin{itemize}

\item A workload-aware QoS scheduling framework (DriftSched) for multi-tenant LLM inference serving on shared GPU infrastructure.

\item A comparative study of workload-characterization fidelity contrasting whitespace-based workload estimation and tokenizer-aware admission-time accounting.

\item An empirical evaluation of FIFO, Priority, Weighted, SJF, and Aging Priority scheduling policies under heterogeneous multi-tenant GPU workloads.

\item An online EMA-based workload calibration mechanism for compensating admission-time workload-estimation error.

\item A reproducible benchmarking and telemetry framework for studying QoS-aware GPU inference scheduling.

\end{itemize}

\subsection{Related Work}
Recent LLM serving systems have focused on improving throughput, memory utilization, and scalability. Nexus introduced scalable GPU cluster scheduling for inference workloads~\cite{nexus2019}. Orca proposed iteration-level scheduling and continuous batching mechanisms that improve accelerator utilization for generative models~\cite{orca2022}. Sarathi further improved inference efficiency through chunked-prefill execution and optimized handling of prefill and decode phases~\cite{sarathi2023}. FlexGen explored high-throughput LLM serving using heterogeneous hardware resources and memory offloading strategies~\cite{flexgen2023}. More recently, vLLM introduced PagedAttention and efficient KV-cache management techniques that significantly improve LLM serving throughput~\cite{vllm2023}.

FastServing explored low-latency distributed inference serving for deep learning workloads and highlighted the importance of efficient request dispatching and scalable serving architectures for production AI systems ~\cite{fastserving2021}.

Recent research has also investigated fairness and isolation mechanisms in multi-tenant LLM serving environments, highlighting the challenges of balancing tenant QoS, resource sharing, and workload interference under shared GPU infrastructure.

Unlike these systems, DriftSched focuses on runtime token drift compensation and adaptive workload estimation for improving admission-time scheduling decisions under multi-tenant GPU contention.

\section{Methodology and System Architecture}

This section describes the proposed adaptive QoS-aware scheduling framework for multi-tenant LLM inference serving on NVIDIA L4 GPUs. The framework was designed to study how workload estimation accuracy, queue management policies, and scheduling algorithms influence fairness, latency, throughput, and GPU utilization under concurrent inference contention. The complete architecture consists of workload generation, workload analysis, tenant-aware queue management, scheduling engines, GPU inference execution, runtime metrics collection, and adaptive feedback-driven workload estimation.

\begin{figure*}[t]
    \centering
    \includegraphics[width=\textwidth]{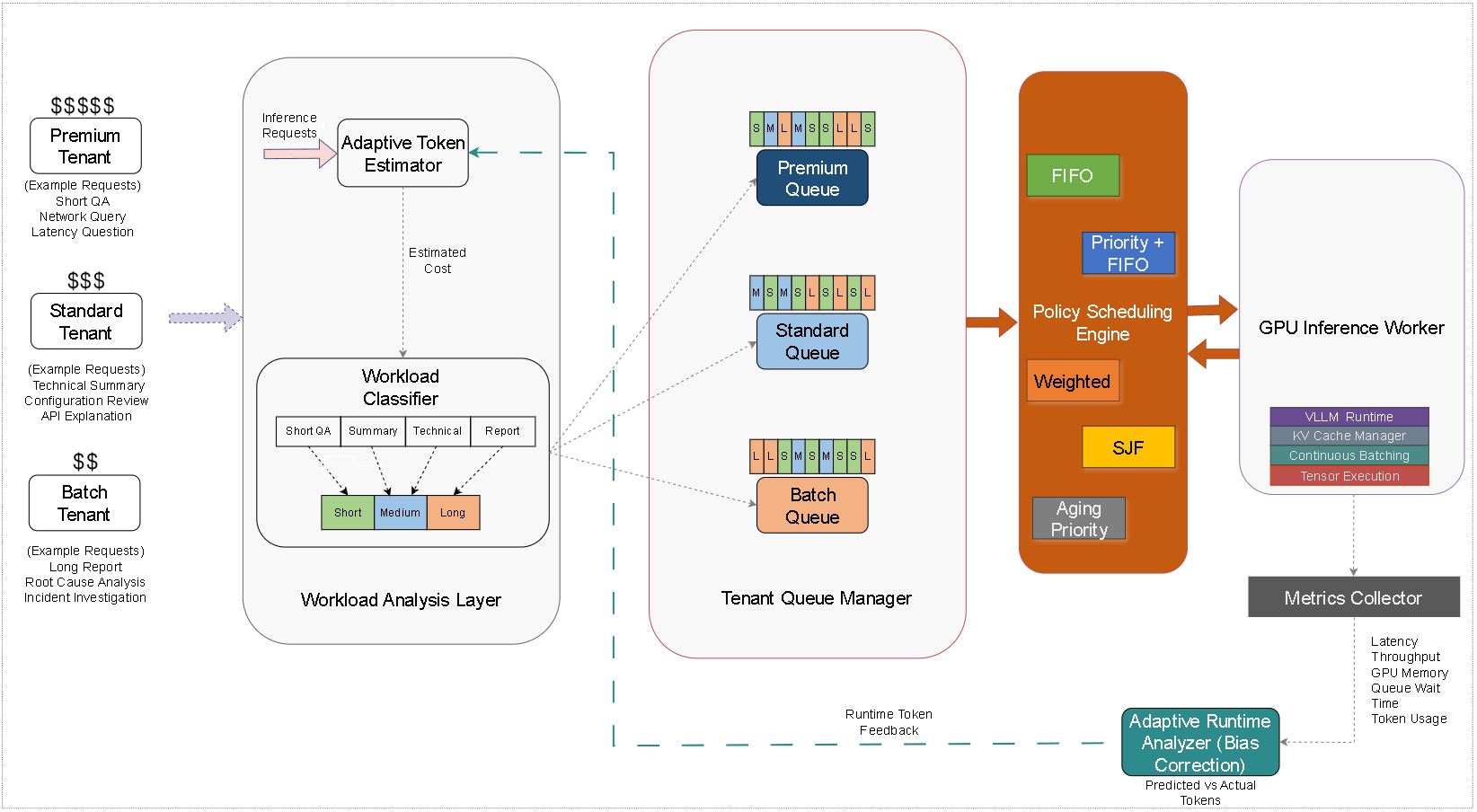}
    \caption{Proposed adaptive QoS-aware multi-tenant LLM inference architecture. Incoming requests are classified using adaptive token-cost estimation and mapped to tenant-specific queues. Multiple scheduling policies (FIFO, Priority Scheduling, Weighted, SJF, and Aging Priority) dispatch requests to the GPU inference worker running vLLM. Runtime metrics are analyzed through a bias-correction feedback loop that continuously refines token-cost estimation based on estimated token budgets versus observed output lengths.}
    \label{fig:qos_architecture}
\end{figure*}

Figure~\ref{fig:qos_architecture} illustrates the overall architecture of the proposed framework.

\subsection{Experimental Workflow}

The experimental workflow begins with the workload generation layer, which produces heterogeneous inference traffic representing multiple tenants and workload categories. Generated requests are submitted to the API Gateway, where the workload characterization layer estimates inference cost, classifies request size, and assigns requests into tenant-specific queues. To evaluate the impact of workload-estimation fidelity, the framework supports two admission-time characterization strategies: a lightweight whitespace-delimited proxy (\texttt{split()}) and tokenizer-aware accounting using the model's native tokenizer.

Following workload characterization, scheduling policies determine the order in which requests are dispatched to the GPU inference worker. Requests are evaluated under FIFO, Priority Scheduling, Weighted Scheduling, Shortest-Job-First (SJF), and Aging Priority Scheduling. Runtime metrics collected during inference execution are used to compare admission-time workload estimates against observed execution behavior. An optional adaptive feedback mechanism applies an Exponential Moving Average (EMA) update rule to maintain workload-specific calibration factors and evaluate the ability of the system to compensate for workload-estimation inaccuracies.

The API Gateway component was implemented using FastAPI, providing lightweight REST-based request submission and integration between workload generation, queue management, and inference execution services~\cite{fastapi}. The framework supports concurrent inference execution under GPU saturation conditions using vLLM-based inference serving on NVIDIA L4 GPUs. Experimental evaluation compares scheduling effectiveness, latency behavior, fairness, queue dynamics, and workload-estimation fidelity under heterogeneous multi-tenant contention scenarios.

\subsection{Workload Analysis Layer}
\label{sec:workload_analysis_layer}

The workload analysis layer estimates request cost before scheduling. It supports both whitespace-based and tokenizer-aware workload characterization and consists of two components: a workload estimator and a workload classifier. Runtime metrics are optionally incorporated through EMA-based calibration to compensate for systematic estimation errors.

\subsubsection{Workload Characterization and Token Budget Estimation}

A major challenge in LLM inference scheduling is accurately characterizing workload cost prior to execution. Scheduling policies such as Shortest-Job-First (SJF), weighted scheduling, admission control, and priority-based queue management rely on admission-time workload estimates to make queue placement and prioritization decisions. Inaccurate workload characterization may propagate directly into scheduling decisions, resulting in request misclassification, queue imbalance, increased tail latency, unfair resource allocation, and degraded Quality-of-Service (QoS) under contention.

The workload estimator supports either whitespace-based or tokenizer-aware accounting, enabling direct comparison between approximate and token-space workload characterization.

\begin{figure}[t]
\centering
\includegraphics[width=0.8\columnwidth]{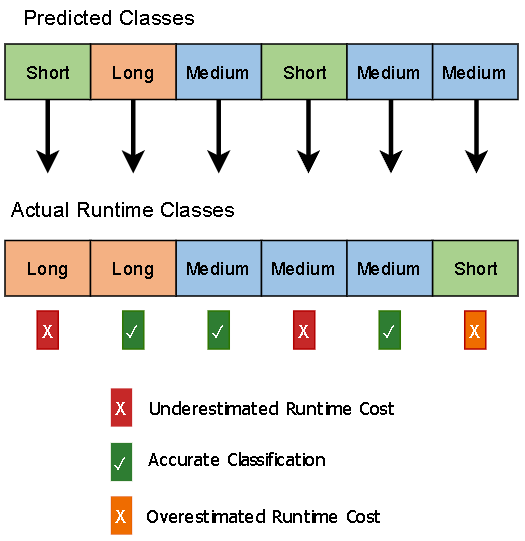}
\caption{Example workload misclassification caused by inaccurate workload characterization. Predicted workload classes may differ from actual runtime classes observed during GPU execution. Underestimation occurs when a request generates more output than expected, while overestimation occurs when actual runtime cost is lower than predicted.}
\label{fig:runtime_misclassification}
\end{figure}

The workload estimator combines workload classification, tenant-aware scaling factors, workload-size estimation, and optional runtime feedback calibration to compute an admission-time token budget prior to GPU execution. Unlike static approaches that rely solely on predefined token limits, DriftSched supports adaptive calibration using execution feedback collected during runtime.

The estimated workload budget is computed as:

\begin{equation}
T_{budget}
=
T_{input}
+
T_{estimated\_output}
\end{equation}

The estimated output token count is calculated using:

\begin{equation}
\begin{aligned}
T_{estimated\_output}
&=
T_{base}
\times
B_{runtime}
\\
&\times
S_{tenant}
\times
F_{input}
\end{aligned}
\end{equation}

where:

\begin{itemize}
\item $T_{base}$ represents the baseline workload token estimate.
\item $B_{runtime}$ represents a runtime calibration factor maintained using execution feedback.
\item $S_{tenant}$ represents tenant-aware safety scaling.
\item $F_{input}$ represents prompt complexity scaling.
\end{itemize}

For the whitespace-proxy configuration, workload size is estimated using whitespace-delimited word counts. For the tokenizer-aware configuration, input and output lengths are measured using the model's native tokenizer and runtime-generated token identifiers. This dual configuration enables direct comparison between approximate and tokenizer-aware workload characterization strategies.

Runtime execution metrics are continuously compared against admission-time estimates to evaluate workload-estimation fidelity. An optional Exponential Moving Average (EMA) feedback mechanism maintains workload-specific calibration factors and enables the framework to compensate for systematic estimation errors when approximate workload characterization is employed.

Algorithm~\ref{alg:adaptive_estimator} illustrates the workload-estimation process.

\begin{algorithm}[htbp]
\caption{Workload Estimation with Optional Runtime Calibration}
\label{alg:adaptive_estimator}

\begin{algorithmic}[1]

\STATE Classify workload category
\STATE Retrieve baseline token estimate
\STATE Retrieve runtime calibration factor
\STATE Determine tenant safety factor
\STATE Compute prompt complexity factor

\STATE Compute estimated output tokens:
\STATE $T_{estimated\_output}
=
T_{base}
\times
B_{runtime}
\times
S_{tenant}
\times
F_{input}$

\STATE Return estimated token budget

\end{algorithmic}
\end{algorithm}

\subsubsection{Workload Classification}

Following workload estimation, requests are classified into scheduling-oriented runtime classes that represent expected computational cost. These runtime classes are utilized by scheduling policies such as Shortest-Job-First (SJF), Weighted Scheduling, and admission-control mechanisms to prioritize requests based on estimated execution complexity.

Classification is performed using the estimated total workload budget:

\begin{equation}
\texttt{short} \leq 128,\quad
128 < \texttt{medium} \leq 512,\quad
\texttt{long} > 512
\end{equation}

The resulting runtime classification provides a lightweight abstraction of expected execution cost and enables scheduling policies to distinguish between latency-sensitive short requests and potentially long-running workloads that may influence queue dynamics and tail latency.

\subsection{Workload Category and Runtime Job Classification}

The proposed framework separates semantic workload classification from runtime scheduling classification. Semantic workload categories describe the logical intent of a request, whereas runtime job classes represent the estimated computational cost utilized by the scheduler.

Requests are initially assigned to one of four semantic workload categories:

\begin{itemize}
    \item \texttt{short\_qa}
    \item \texttt{summary}
    \item \texttt{technical}
    \item \texttt{report}
\end{itemize}

Each category is associated with a baseline workload estimate that contributes to admission-time workload characterization. The estimated workload budget is subsequently mapped into one of three runtime scheduling classes:

\[
\texttt{short}, \quad
\texttt{medium}, \quad
\texttt{long}
\]

using the following classification rule:

\begin{equation}
\texttt{job\_type}
=
\begin{cases}
\texttt{short}, & T_{budget} \leq 128 \\
\texttt{medium}, & 128 < T_{budget} \leq 512 \\
\texttt{long}, & T_{budget} > 512
\end{cases}
\end{equation}

In the current implementation, \texttt{short\_qa} workloads frequently map to short runtime jobs, while \texttt{summary} and \texttt{technical} workloads commonly map to medium runtime jobs. Long-form \texttt{report} workloads typically map to medium or long runtime classes due to their larger estimated workload budgets. However, the final runtime classification is determined by the estimated workload budget rather than the semantic category itself.

Runtime scheduling decisions depend on estimated workload cost rather than semantic category alone. Both semantic labels and runtime classes are recorded for subsequent analysis.

\begin{table}[h]
\centering
\caption{Example Mapping Between Semantic Workload Categories and Runtime Scheduling Classes}
\label{tab:workload_mapping}
\small
\begin{tabular}{|p{2.2cm}|p{2.2cm}|p{2.8cm}|}
\hline
\textbf{Semantic Category} & \textbf{Typical Runtime Class} & \textbf{Example Prompt} \\
\hline

\texttt{short\_qa}
&
\texttt{short}
&
``What is DNS?'' \\
\hline

\texttt{summary}
&
\texttt{medium}
&
``Summarize how Kubernetes schedules pods.'' \\
\hline

\texttt{technical}
&
\texttt{medium / long}
&
``Explain distributed systems consistency and retries.'' \\
\hline

\texttt{report}
&
\texttt{medium / long}
&
``Write a detailed report on network outage summarizing affected services.'' \\
\hline

\end{tabular}
\end{table}

\subsection{Tenant Queue Management}

After workload characterization, requests are grouped into tenant-specific service queues managed by the tenant queue manager. Three independent queues are maintained:

\begin{itemize}
    \item Premium Queue
    \item Standard Queue
    \item Batch Queue
\end{itemize}

Each queue stores heterogeneous workloads consisting of short, medium, and long runtime classes. Tenant isolation enables scheduling policies to enforce QoS differentiation under concurrent contention while preserving workload-awareness during request selection.

The queue manager supports multiple queue implementations using Redis-based distributed data structures~\cite{redis}. FIFO scheduling utilizes Redis lists to preserve request arrival order. Priority, Weighted, and Aging Priority schedulers maintain tenant-specific queues and apply scheduling policies during request selection. SJF scheduling additionally incorporates runtime workload classification to prioritize requests with lower estimated execution cost.

Because scheduling policies operate directly on workload-characterization outputs, workload-estimation fidelity has a direct influence on queue composition, scheduler decisions, fairness, and latency behavior.

\subsection{Scheduling Engine}

The scheduling engine determines the order in which queued inference requests are dispatched to the GPU inference worker. Multiple scheduling strategies were implemented to evaluate fairness, latency behavior, starvation characteristics, and throughput under contention.

The evaluated scheduling policies are inspired by traditional operating system scheduling techniques and large-scale distributed scheduling systems such as Sparrow~\cite{sparrow}, as well as modern multi-tenant accelerator scheduling architectures~\cite{nexus2019}. The objective is to study how workload-characterization fidelity influences queue behavior, scheduler effectiveness, and tenant QoS under GPU saturation.

\subsection{Calibration and Evaluation Phases}

To ensure consistent comparison across scheduling policies and workload-characterization strategies, each experimental run was divided into a calibration phase and a stress evaluation phase. A total of 3000 inference requests were generated for each run.

Each experiment consisted of a 1000-request calibration phase followed by a 2000-request stress phase (1:2 ratio). Runtime feedback remained enabled throughout execution.

Unlike offline training approaches, DriftSched performs continuous online adaptation throughout execution. Runtime observations collected during both phases are incorporated into subsequent workload estimates through EMA-based calibration updates. This enables the framework to evaluate scheduler behavior under both approximate workload-characterization scenarios and tokenizer-aware accounting while maintaining a consistent experimental methodology across all evaluated configurations.

\begin{figure}[t]
\centering
\includegraphics[width=\columnwidth]{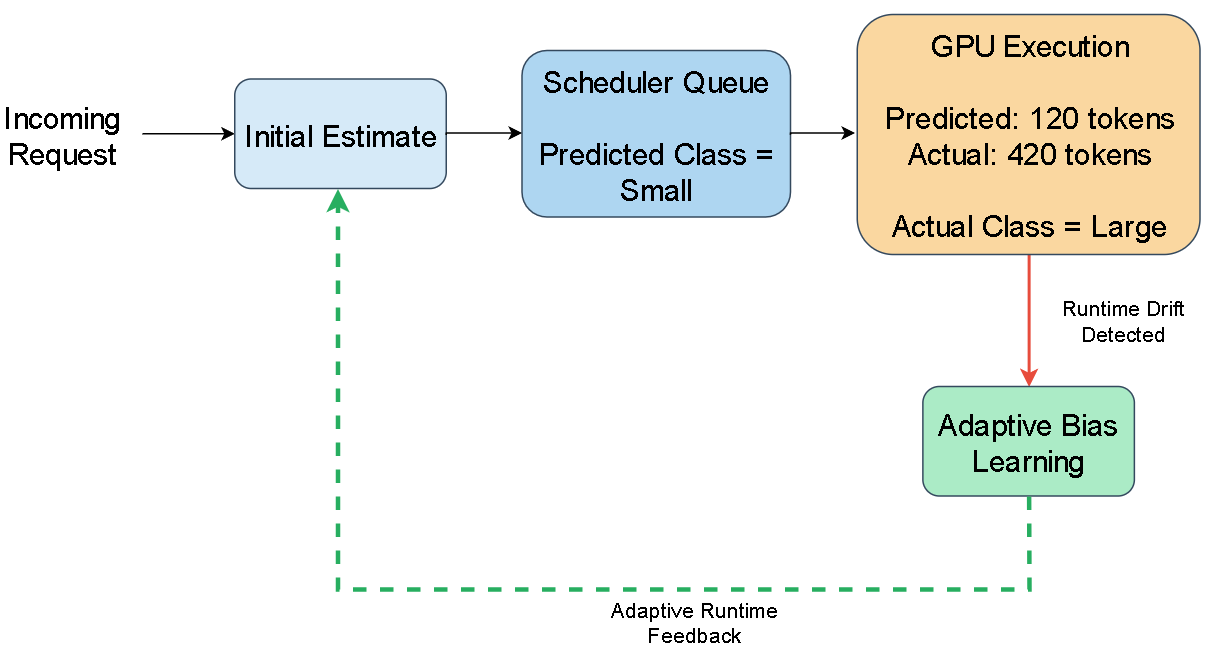}
\caption{DriftSched adaptive runtime learning mechanism. Runtime token drift is detected by comparing estimated token budgets and observed output lengths during GPU execution. The observed prediction error is used to update workload-specific bias factors, which are subsequently fed back into the workload estimation layer to improve future scheduling decisions.}
\label{fig:driftsched_feedback_loop}
\end{figure}

\subsubsection{FIFO Scheduling}

FIFO scheduling dispatches requests strictly in arrival order~\cite{osconcepts} without considering workload size or tenant priority. While FIFO provides fairness in terms of arrival ordering, long-running requests may block smaller latency-sensitive requests under burst traffic conditions.

\begin{table*}[t]
\centering
\caption{Experimental Platform Configuration}
\label{tab:experimental_platform}

\begin{subtable}[t]{0.48\textwidth}
\centering
\caption{Host System Configuration}
\label{tab:hardware_specs}
\small
\begin{tabular}{p{3.0cm} p{4.0cm}}
\hline
\textbf{Component} & \textbf{Specification} \\
\hline
CPU & Intel Xeon 6 Performance (Granite Rapids), 24 cores / 48 threads \\
Memory & 32\,GB DDR5-6400 \\
Storage & NVMe SSD \\
Operating System & Ubuntu 22.04 LTS \\
GPU & NVIDIA L4 \\
GPU Memory & 24\,GB GDDR6 \\
GPU Interface & PCIe Gen4 x16 \\
GPU Power Limit & 72\,W \\
NVIDIA Driver & 535.288.01 \\
CUDA Version & CUDA 12.1 \\
GPU Telemetry & NVIDIA-SMI \\
\hline
\end{tabular}
\end{subtable}
\hfill
\begin{subtable}[t]{0.48\textwidth}
\centering
\caption{Software Environment Configuration}
\label{tab:software_specs}
\small
\begin{tabular}{p{3.0cm} p{4.0cm}}
\hline
\textbf{Software Component} & \textbf{Specification} \\
\hline
Programming Language & Python 3.10.12 \\
PyTorch Version & PyTorch 2.4.0 + CUDA 12.1 \\
Inference Runtime & vLLM (development build) \\
LLM Model & Qwen1.5-1.8B-Chat \\
Model Precision & FP16 \\
API Framework & FastAPI \\
Queue Backend & Redis \\
Concurrency Framework & Python ThreadPoolExecutor \\
Metrics Storage & CSV Logging \\
GPU Monitoring Interval & 200\,ms \\
\hline
\end{tabular}
\end{subtable}

\end{table*}

\subsubsection{Priority Scheduling}

Priority Scheduling assigns higher execution precedence to Premium tenants while preserving FIFO ordering within each priority tier. Separate queues are maintained for Premium, Standard, and Batch tenants. During scheduling, requests are selected from the highest-priority non-empty queue, ensuring that Premium workloads receive preferential access to GPU resources. Requests within the same tenant tier are processed in arrival order using FIFO semantics. The scheduling score is computed as: 
\[ 
\texttt{score} 
= 
(\texttt{priority\_score} \times 10^{12}) 
+ 
\texttt{arrival\_time} 
\] 
This mechanism ensures: 
\begin{itemize} 
\item Premium requests execute before Standard requests. 
\item Standard requests execute before Batch requests. 
\item FIFO ordering is preserved within each priority tier. 
\end{itemize}

\subsubsection{Shortest-Job-First Scheduling}

SJF scheduling~\cite{osconcepts,queueing} prioritizes requests with lower estimated workload budgets. The scheduler utilizes workload-characterization outputs produced by the workload analysis layer to prioritize requests with lower expected execution cost. As a result, SJF is particularly sensitive to workload-estimation fidelity and provides a useful evaluation platform for studying how admission-time workload characterization influences latency and queue behavior.

\subsubsection{Weighted Scheduling}

Weighted scheduling~\cite{sparrow} partitions GPU service capacity across tenant classes using predefined execution ratios. The proposed implementation uses a

\[
50/30/20
\]

distribution for

\[
\texttt{Premium} :
\texttt{Standard} :
\texttt{Batch}
\]

requests, respectively. The scheduler cyclically dispatches requests from tenant-specific queues according to the configured ratio. Similar proportional-share resource allocation mechanisms have been explored in operating systems through lottery scheduling~\cite{lottery}, where service shares are allocated according to configurable weights.

\subsubsection{Aging Priority Scheduling}

To mitigate starvation~\cite{osconcepts}, Aging Priority Scheduling dynamically increases request priority as queue waiting time grows. Waiting time progressively reduces the effective scheduling score, allowing long-waiting requests to eventually execute even under continuous high-priority traffic.

\subsection{GPU Inference Execution}

Inference execution is performed using vLLM~\cite{vllm, vllm2023} running on NVIDIA L4 GPUs. The runtime leverages PagedAttention-based KV-cache management and continuous batching mechanisms originally proposed for efficient LLM serving. The GPU inference worker loads the Qwen1.5-1.8B-Chat model using FP16 precision for optimized inference throughput. Model execution was performed using the PyTorch deep learning framework operating in FP16 precision mode through the vLLM inference runtime environment~\cite{pytorch}. The inference runtime leverages several vLLM optimizations including continuous batching, KV-cache management, optimized tensor execution, and GPU memory-aware scheduling. Inference requests are executed using configurable sampling parameters including temperature and maximum token generation limits.

\subsection{Runtime Metrics Collection}

Runtime metrics are collected throughout the inference lifecycle to evaluate scheduling behavior under contention. Metrics include queue wait time, inference latency, end-to-end latency, observed output length, GPU memory utilization, GPU utilization, throughput, and scheduling fairness. Worker execution timestamps are recorded before and after GPU execution to separate queueing latency from inference runtime latency. Metrics are persisted into CSV files for post-processing and visualization. GPU telemetry collection is performed using \texttt{nvidia-smi} sampling at 200 millisecond intervals during experimental execution.

\subsection{Runtime Calibration and Feedback Adaptation}

DriftSched incorporates an optional runtime calibration mechanism that continuously compares admission-time workload estimates against observed execution behavior. Runtime feedback is used to refine workload estimates when characterization errors are present.

During inference execution, observed output length is compared against the workload estimate generated during admission-time analysis. The resulting estimation error is incorporated into workload-specific calibration factors using an Exponential Moving Average (EMA) update rule. These calibration factors are subsequently utilized by future workload-estimation decisions.

The adaptive update rule is computed as:

\begin{equation}
B_{new}
=
(1 - \alpha)
\times
B_{old}
+
\alpha
\times
B_{measured}
\end{equation}

where:

\begin{equation}
B_{measured}
=
\frac{
T_{actual}
}{
T_{base}
}
\end{equation}

where $B_{new}$ represents the updated calibration factor, $B_{old}$ represents the previous calibration factor, $\alpha$ represents the EMA learning rate, $B_{measured}$ represents the observed calibration ratio, $T_{actual}$ represents the observed output length, and $T_{base}$ represents the baseline workload estimate associated with the workload category.

EMA smoothing incorporates runtime observations while limiting sensitivity to transient fluctuations.

Runtime feedback enables DriftSched to operate under both approximate and tokenizer-aware workload-characterization strategies using a common estimation framework.

\begin{figure*}[t]
\centering

\begin{subfigure}{0.32\textwidth}
\centering
\includegraphics[width=\linewidth]{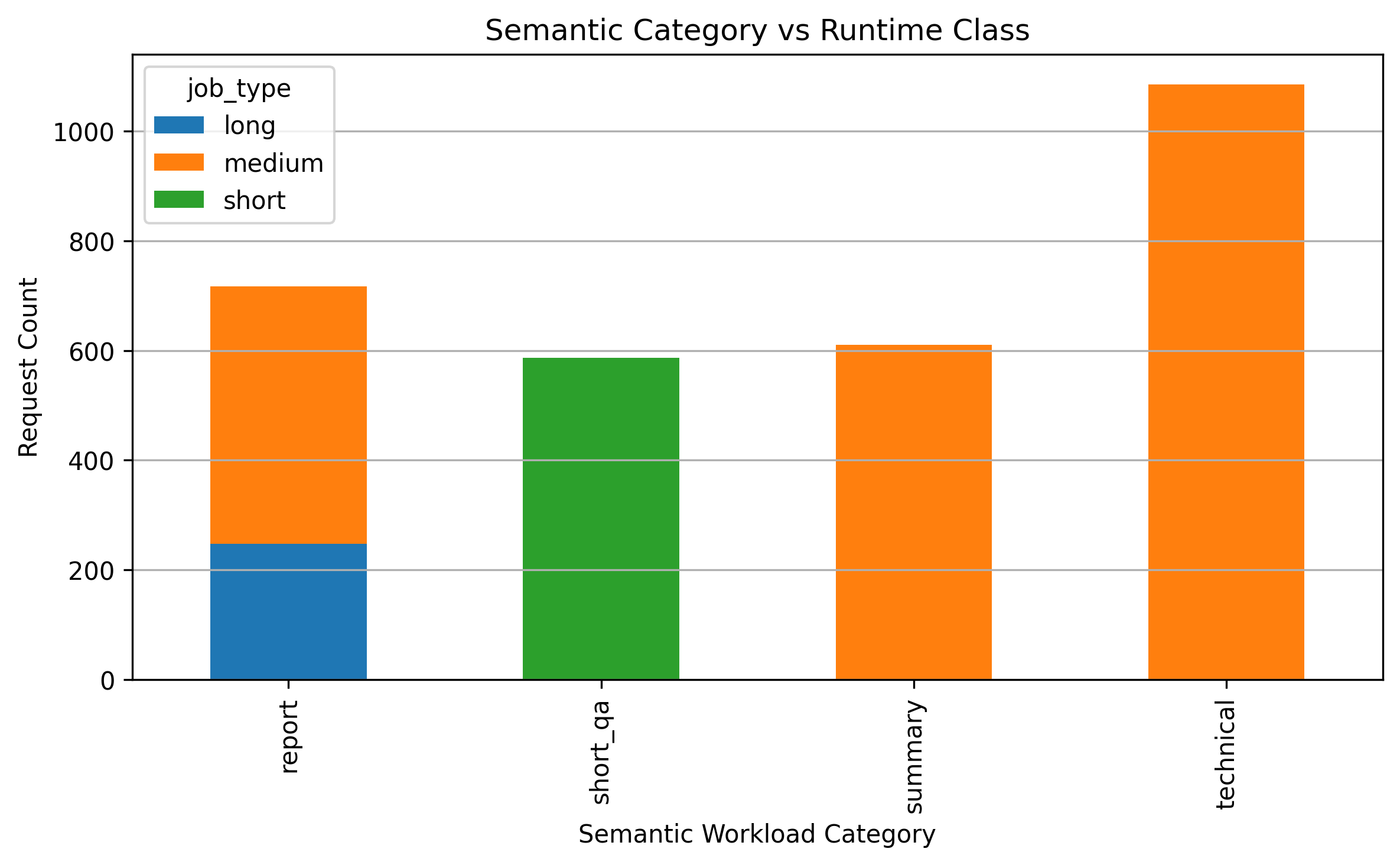}
\caption{FIFO}
\end{subfigure}
\hfill
\begin{subfigure}{0.32\textwidth}
\centering
\includegraphics[width=\linewidth]{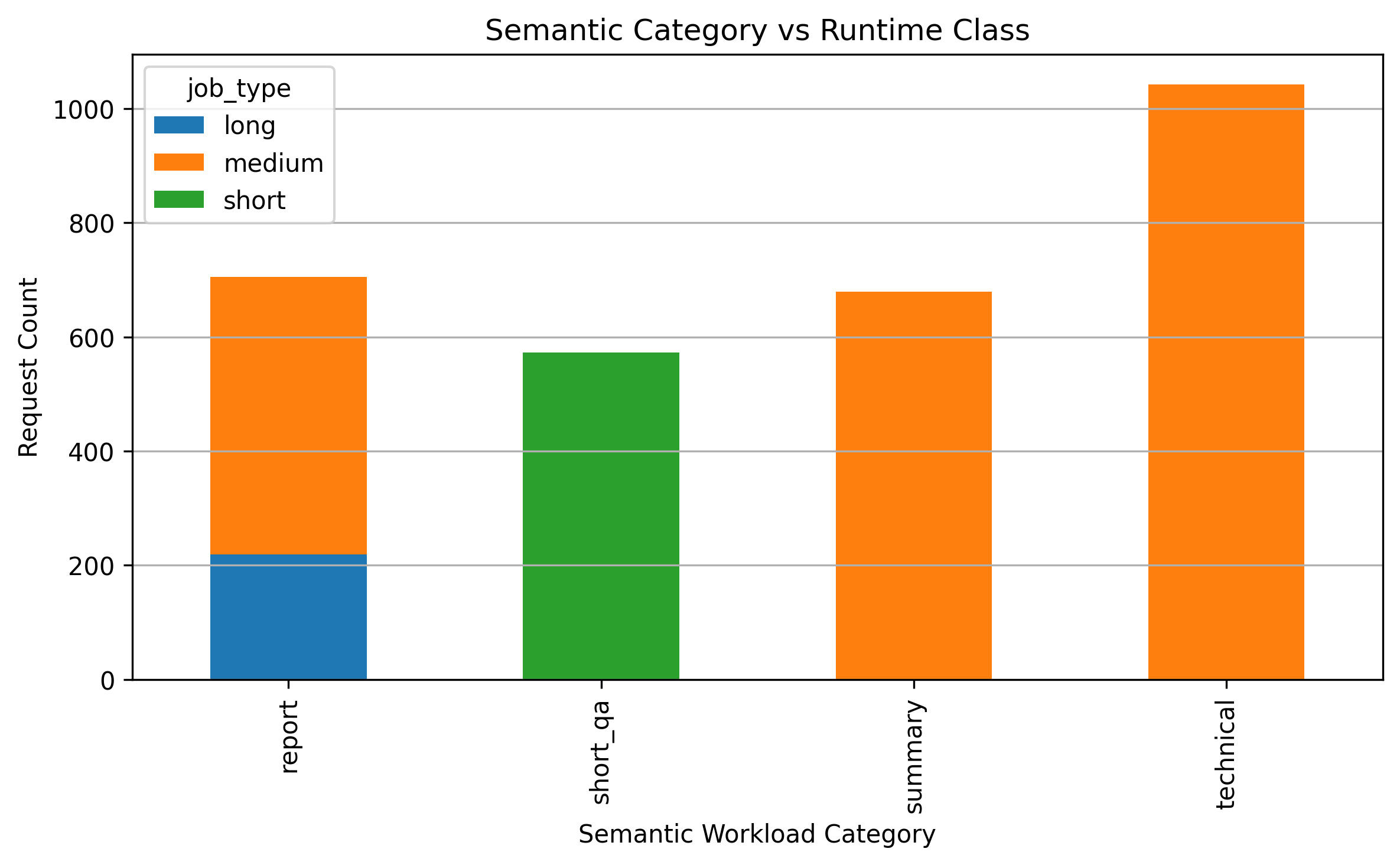}
\caption{Priority }
\end{subfigure}
\hfill
\begin{subfigure}{0.32\textwidth}
\centering
\includegraphics[width=\linewidth]{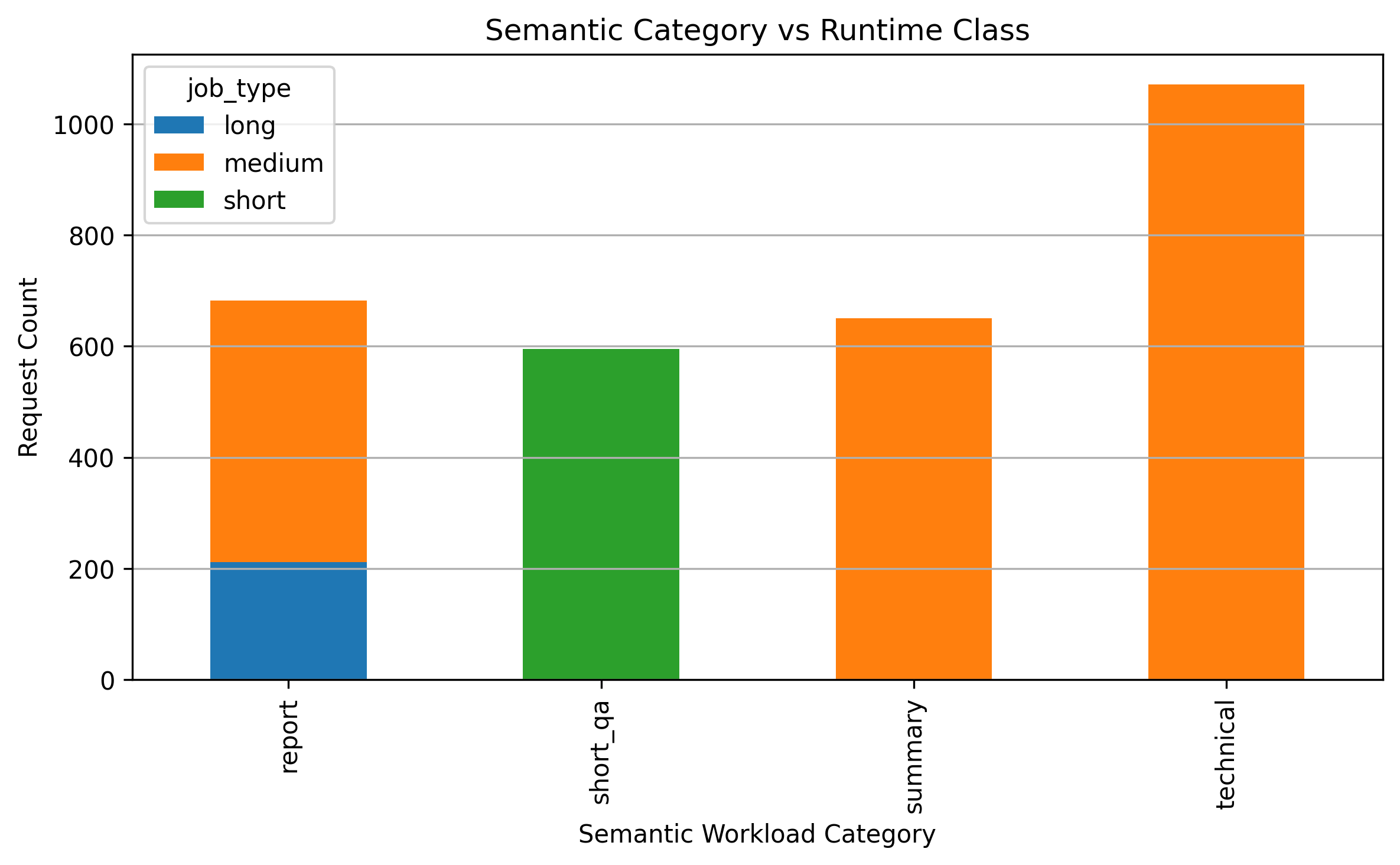}
\caption{Weighted}
\end{subfigure}
\vspace{0.5em}
\makebox[\textwidth][c]{%
\begin{subfigure}{0.32\textwidth}
\centering
\includegraphics[width=\linewidth]{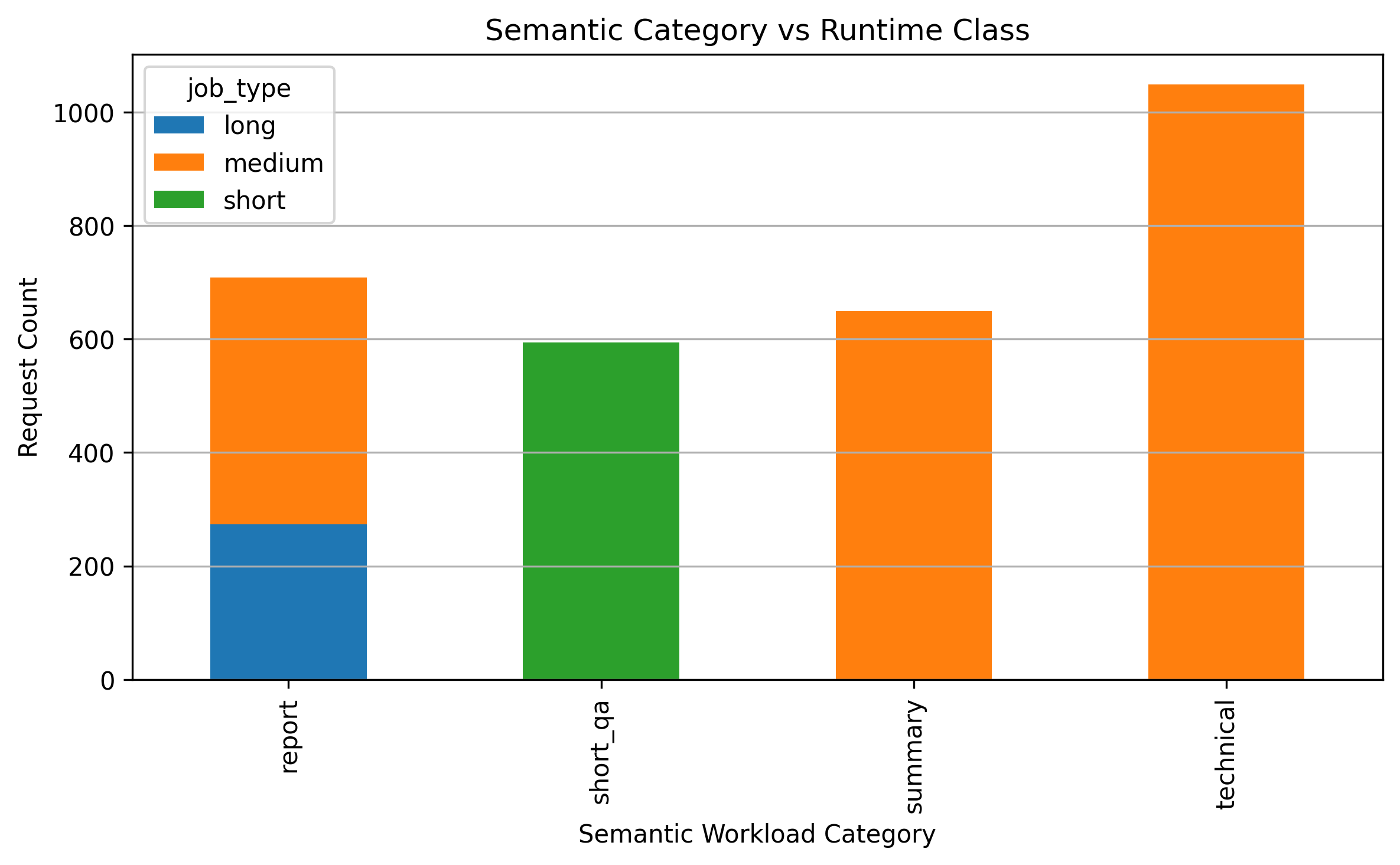}
\caption{SJF}
\end{subfigure}
\hspace{0.04\textwidth}
\begin{subfigure}{0.32\textwidth}
\centering
\includegraphics[width=\linewidth]{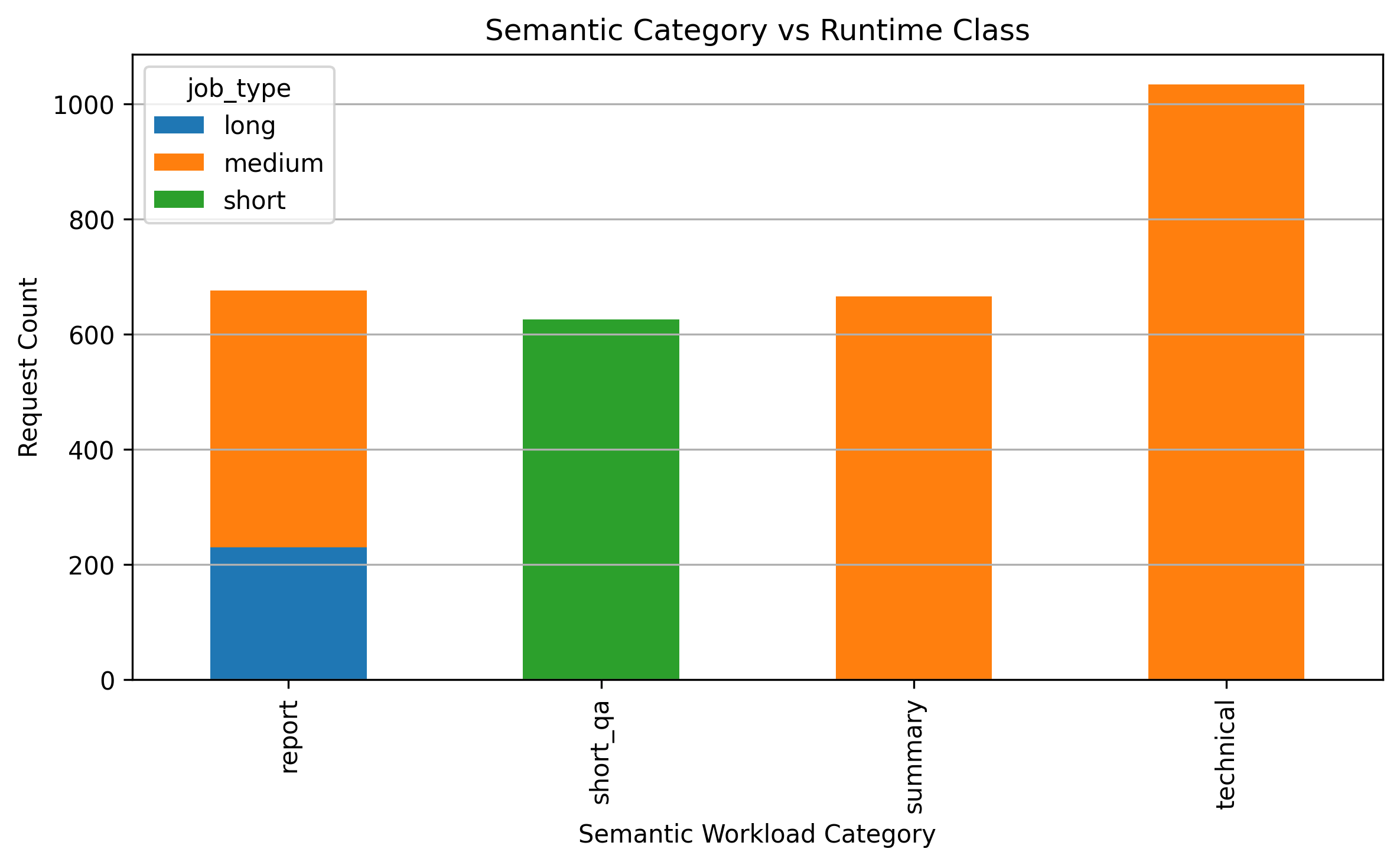}
\caption{Aging Priority}
\end{subfigure}
}
\caption{
Semantic workload categories versus runtime scheduling classes using whitespace-based workload estimation (\texttt{split()}). Report-generation workloads span both medium and long runtime classes, while technical and summarization workloads are predominantly classified as medium jobs. These results illustrate how coarse workload-characterization heuristics can influence admission-time scheduling decisions.
}
\label{fig:semantic_vs_runtime_classification}
\end{figure*}

\begin{figure*}[t]
\centering

\begin{subfigure}{0.32\textwidth}
\centering
\includegraphics[width=\linewidth]{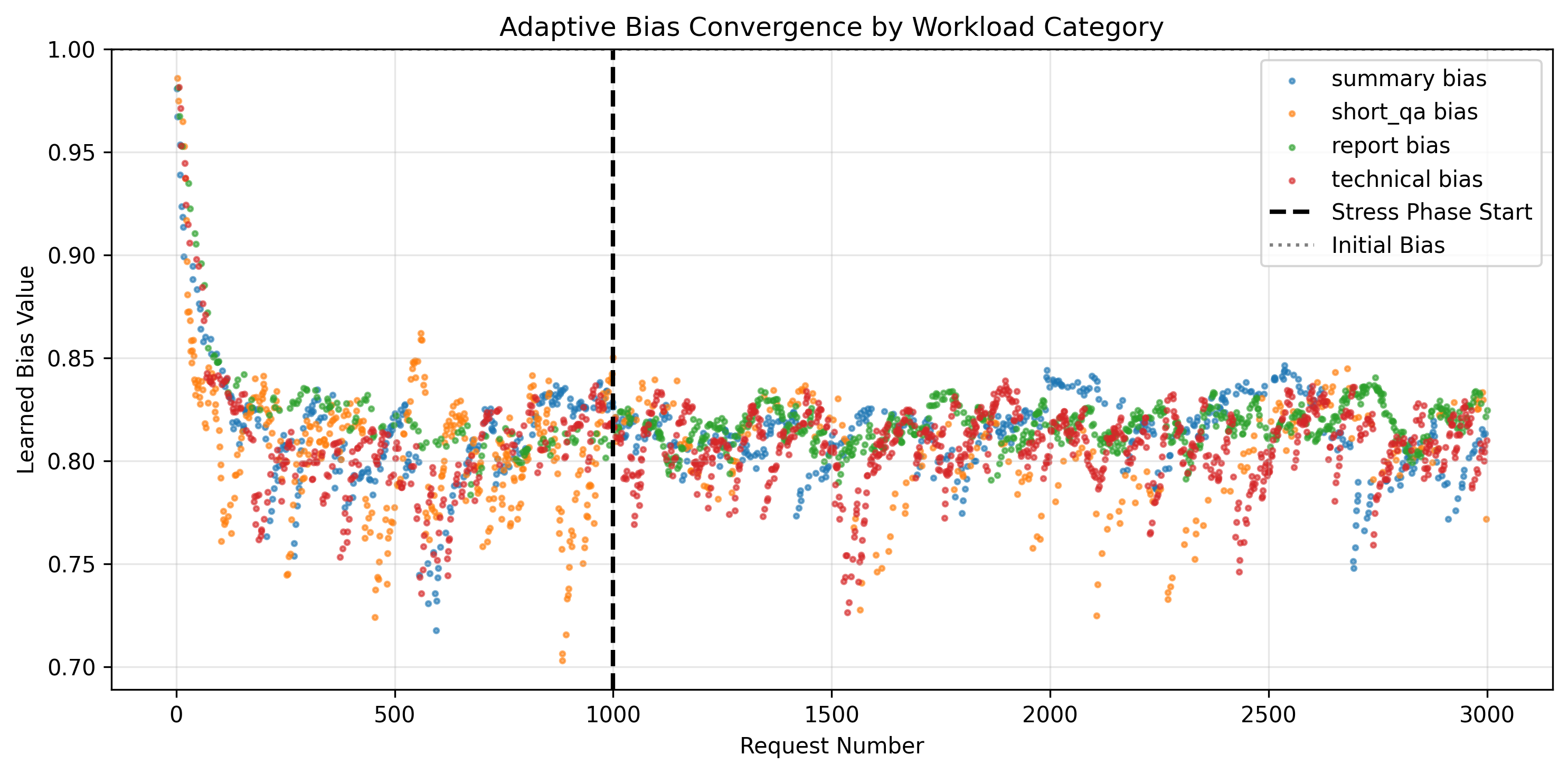}
\caption{FIFO}
\end{subfigure}
\hfill
\begin{subfigure}{0.32\textwidth}
\centering
\includegraphics[width=\linewidth]{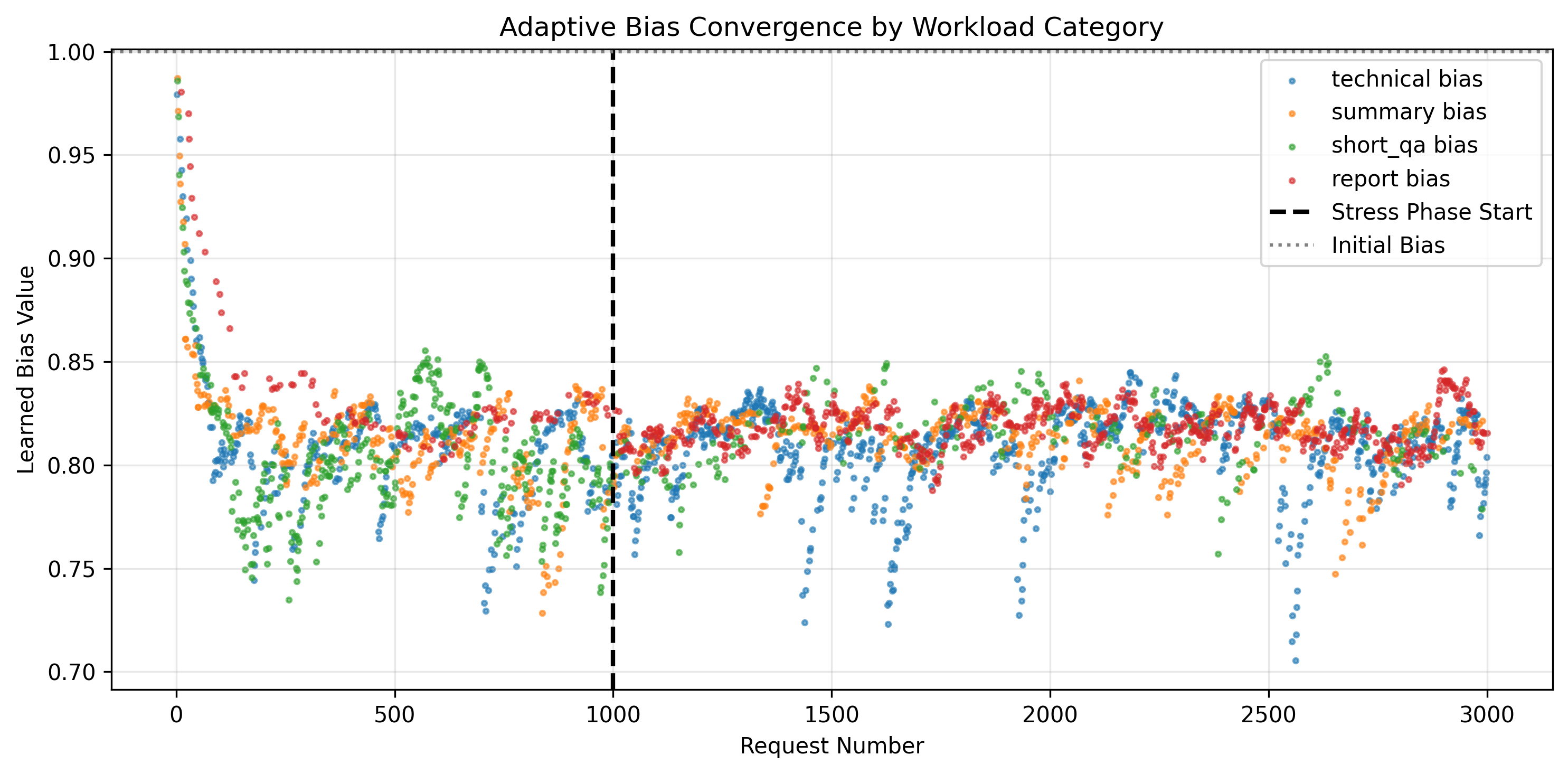}
\caption{Priority }
\end{subfigure}
\hfill
\begin{subfigure}{0.32\textwidth}
\centering
\includegraphics[width=\linewidth]{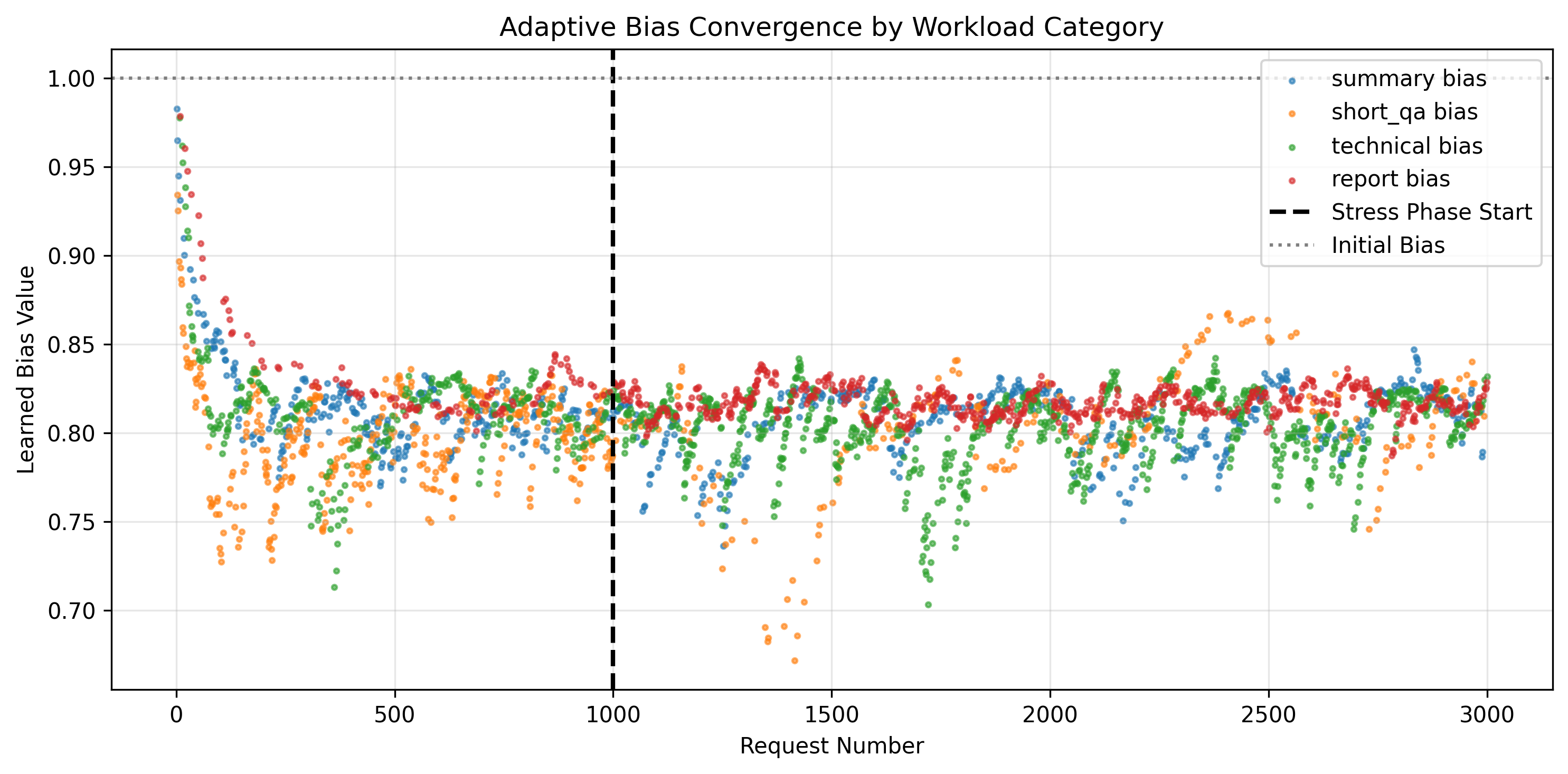}
\caption{Weighted}
\end{subfigure}

\vspace{0.5em}

\makebox[\textwidth][c]{%
\begin{subfigure}{0.38\textwidth}
\centering
\includegraphics[width=\linewidth]{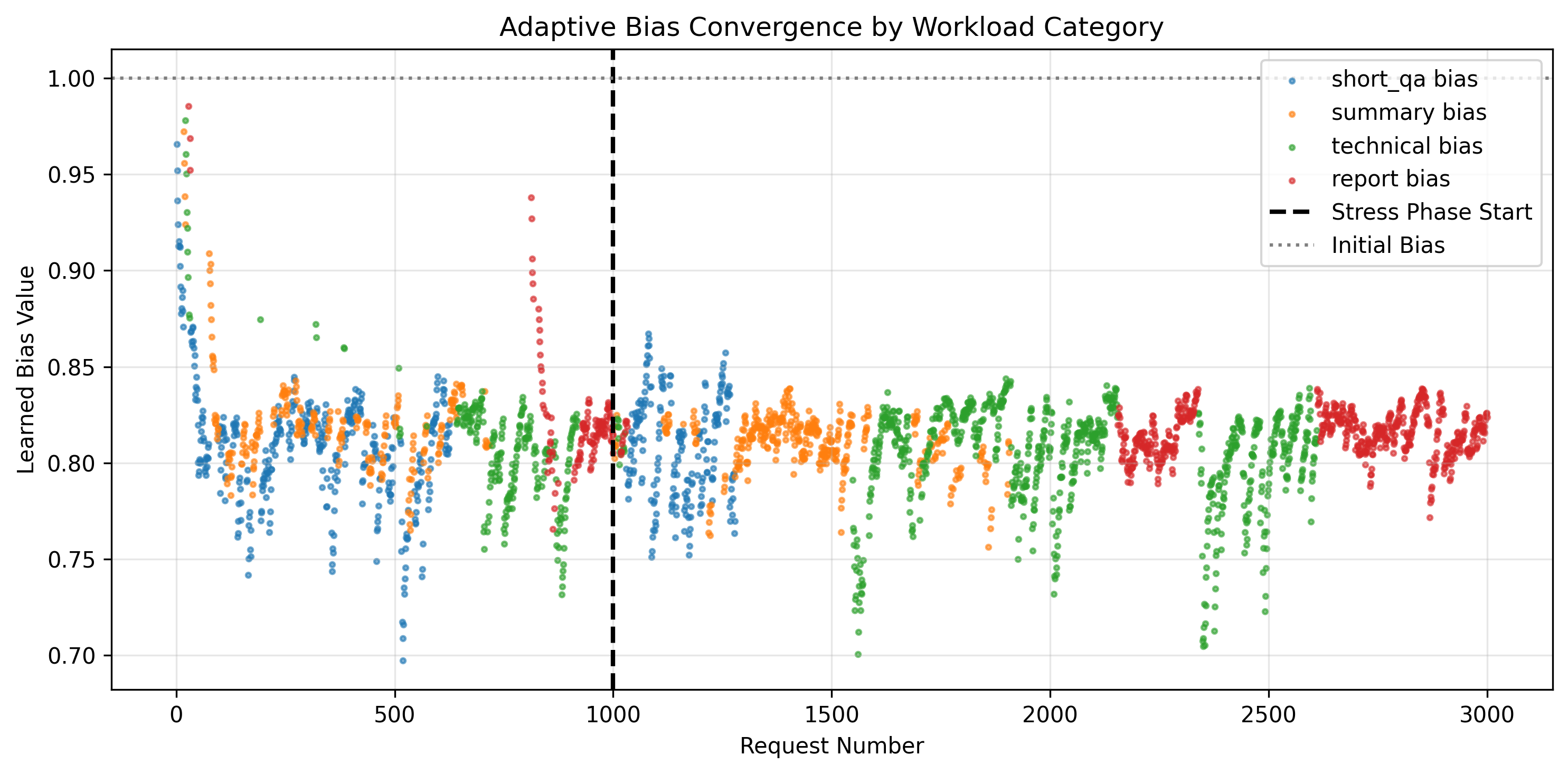}
\caption{SJF}
\end{subfigure}
\hspace{0.04\textwidth}
\begin{subfigure}{0.38\textwidth}
\centering
\includegraphics[width=\linewidth]{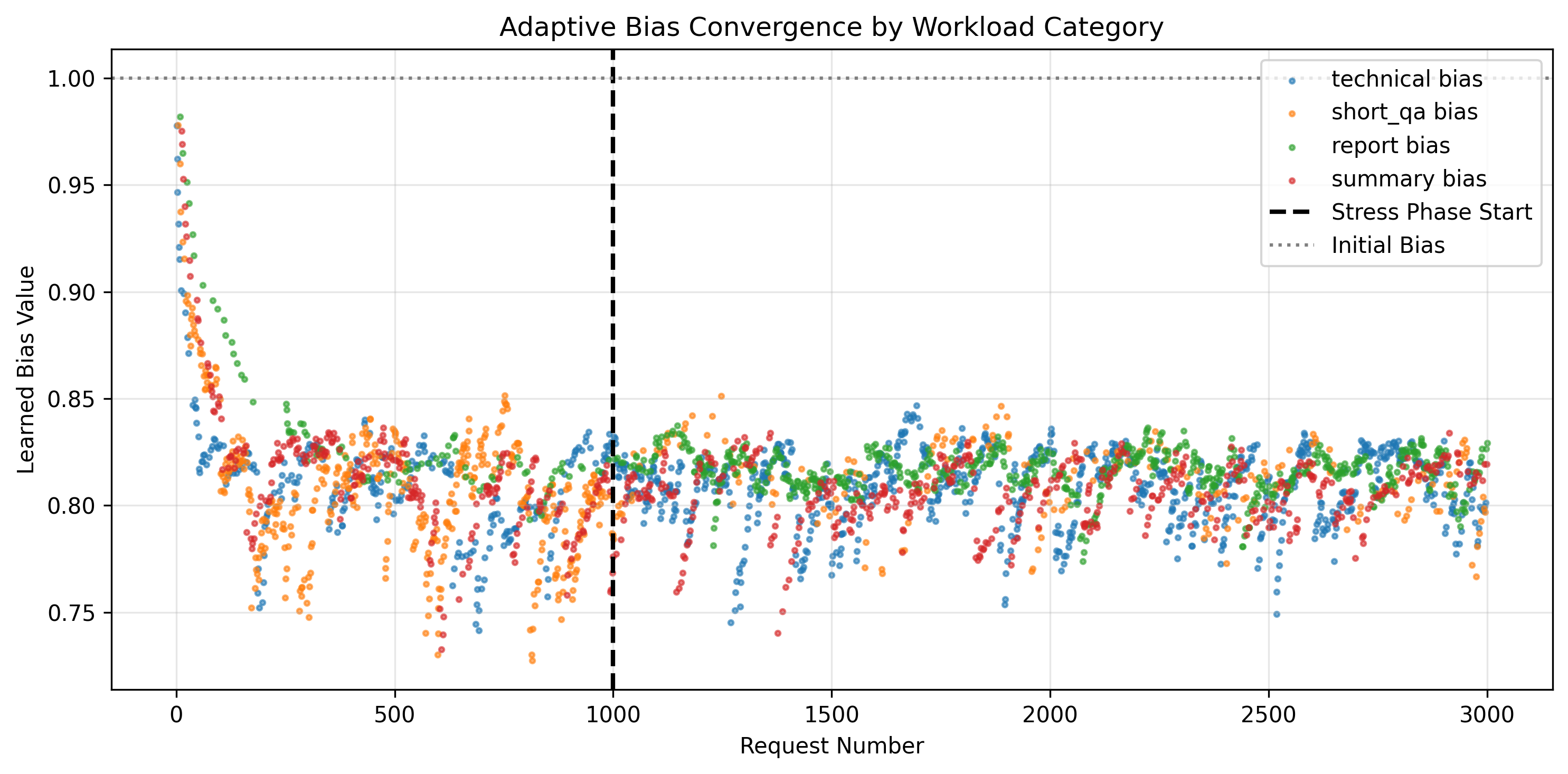}
\caption{Aging Priority}
\end{subfigure}
}

\caption{
Adaptive bias convergence under whitespace-based workload characterization (\texttt{split()} + BIAS=ON) for FIFO, Priority, Weighted, SJF, and Aging Priority scheduling. Bias factors are initialized to 1.0 and updated using EMA-based runtime feedback. The dashed line marks the transition from the calibration phase (first 1000 requests) to the stress phase. Across all schedulers, bias values converge to approximately 0.79--0.84, indicating systematic estimation error introduced by whitespace-based workload characterization.
}
\label{fig:bias_convergence_all_schedulers}
\end{figure*}

\section{Experimental Setup}
\label{sec:experimental_methodology}
\subsection{Experimental Environment}

\begin{figure*}[t]
\centering

\begin{subfigure}{0.32\textwidth}
\centering
\includegraphics[width=\linewidth]{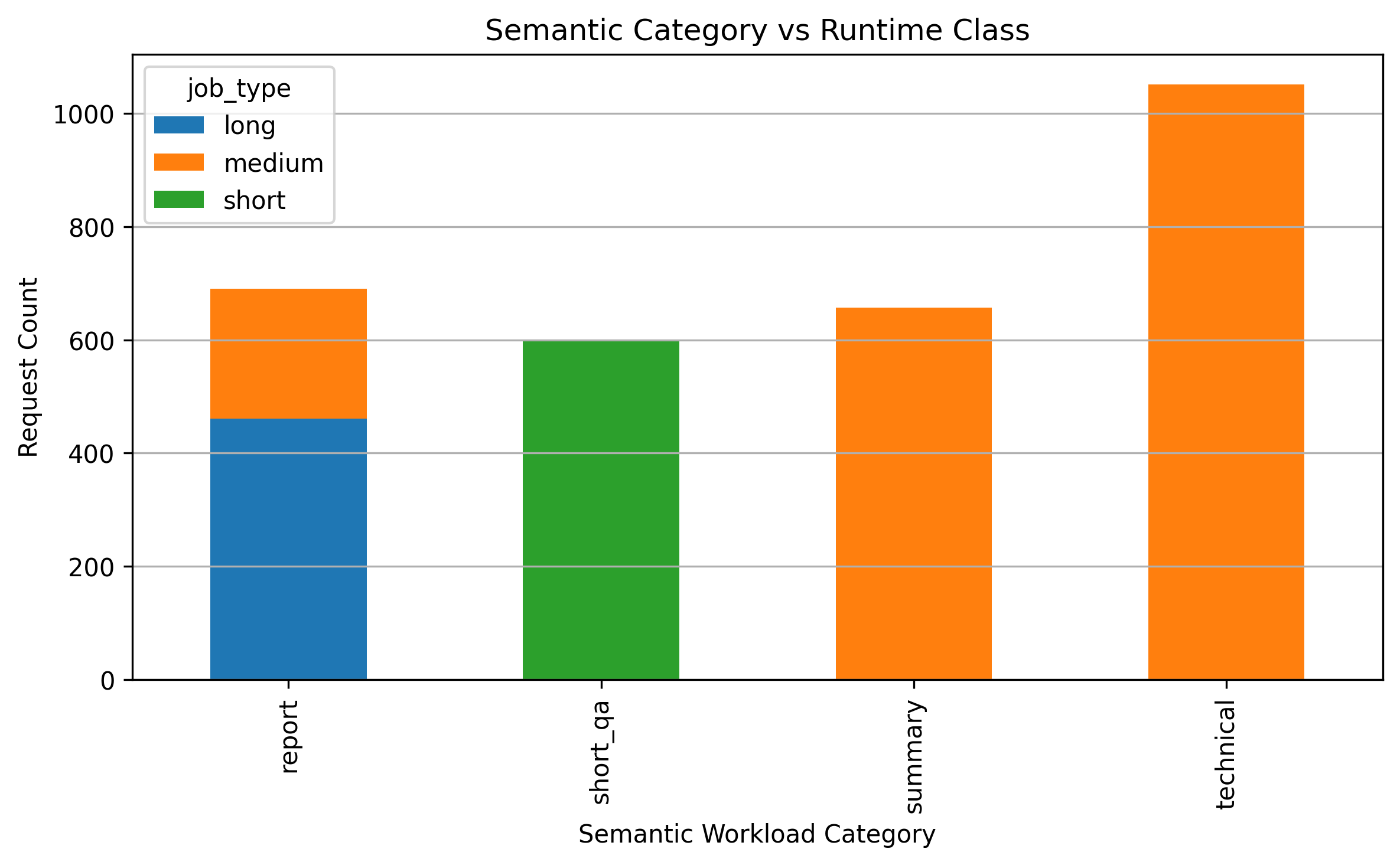}
\caption{FIFO}
\end{subfigure}
\hfill
\begin{subfigure}{0.32\textwidth}
\centering
\includegraphics[width=\linewidth]{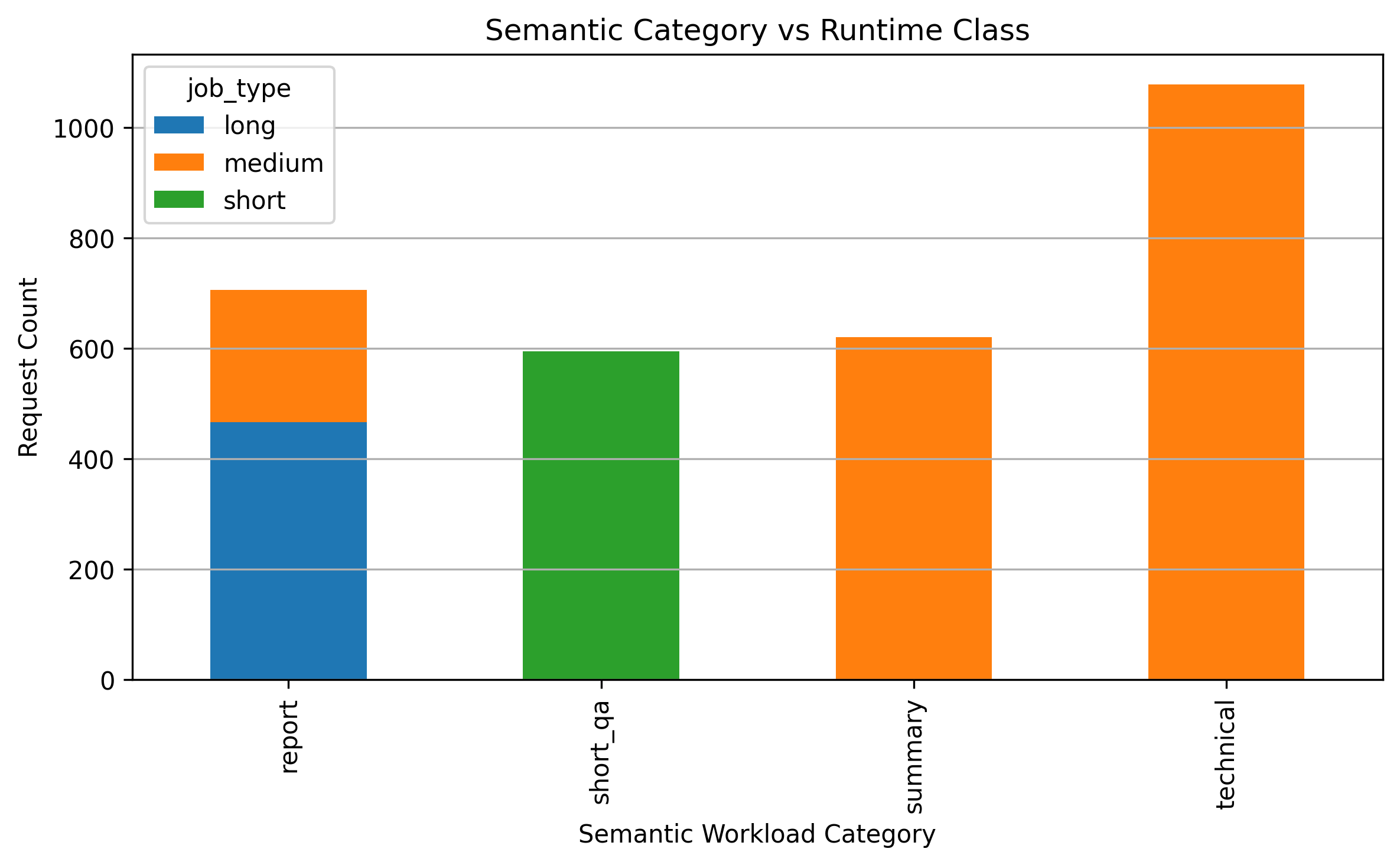}
\caption{Priority}
\end{subfigure}
\hfill
\begin{subfigure}{0.32\textwidth}
\centering
\includegraphics[width=\linewidth]{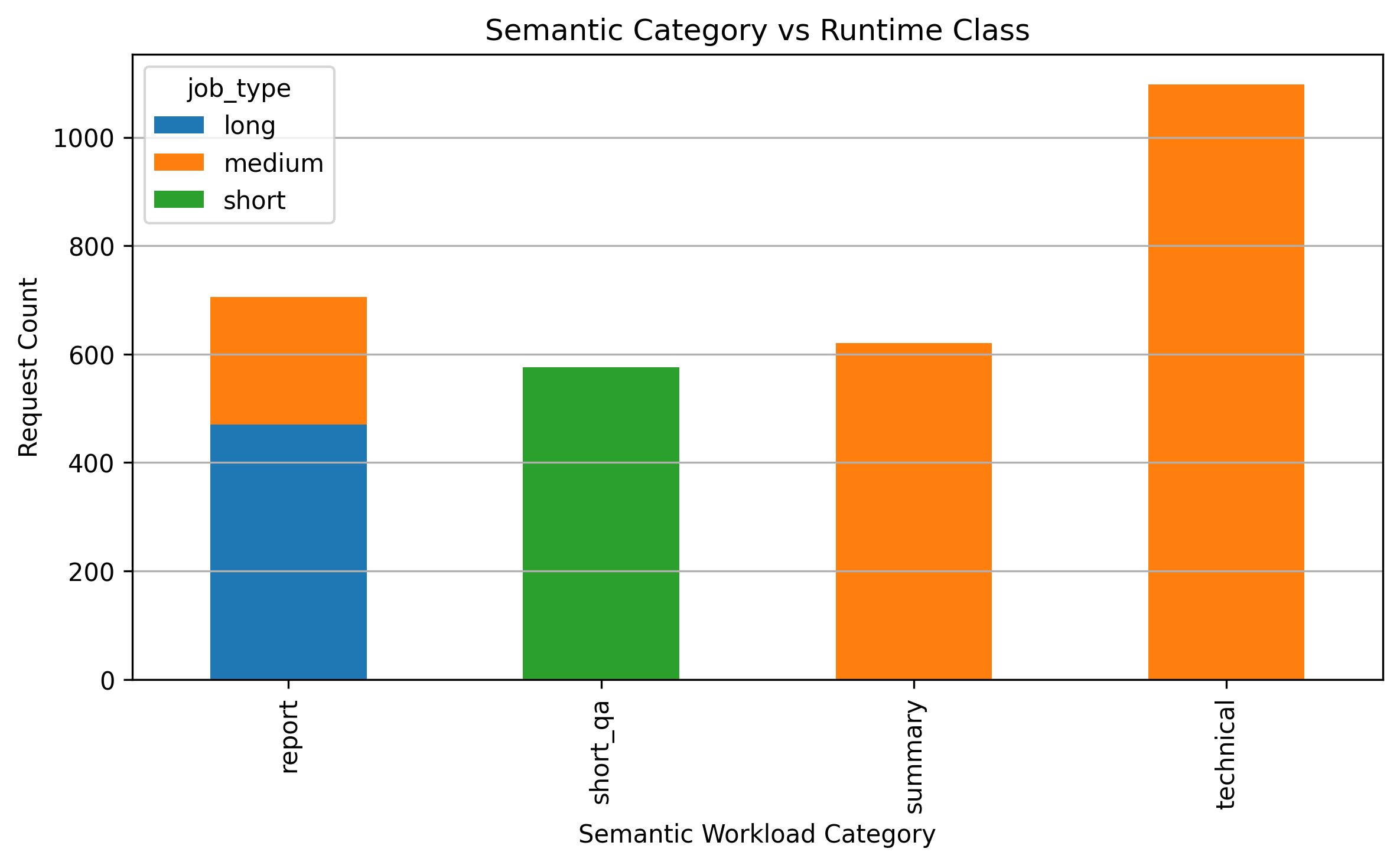}
\caption{Weighted}
\end{subfigure}

\vspace{0.5em}

\makebox[\textwidth][c]{%
\begin{subfigure}{0.32\textwidth}
\centering
\includegraphics[width=\linewidth]{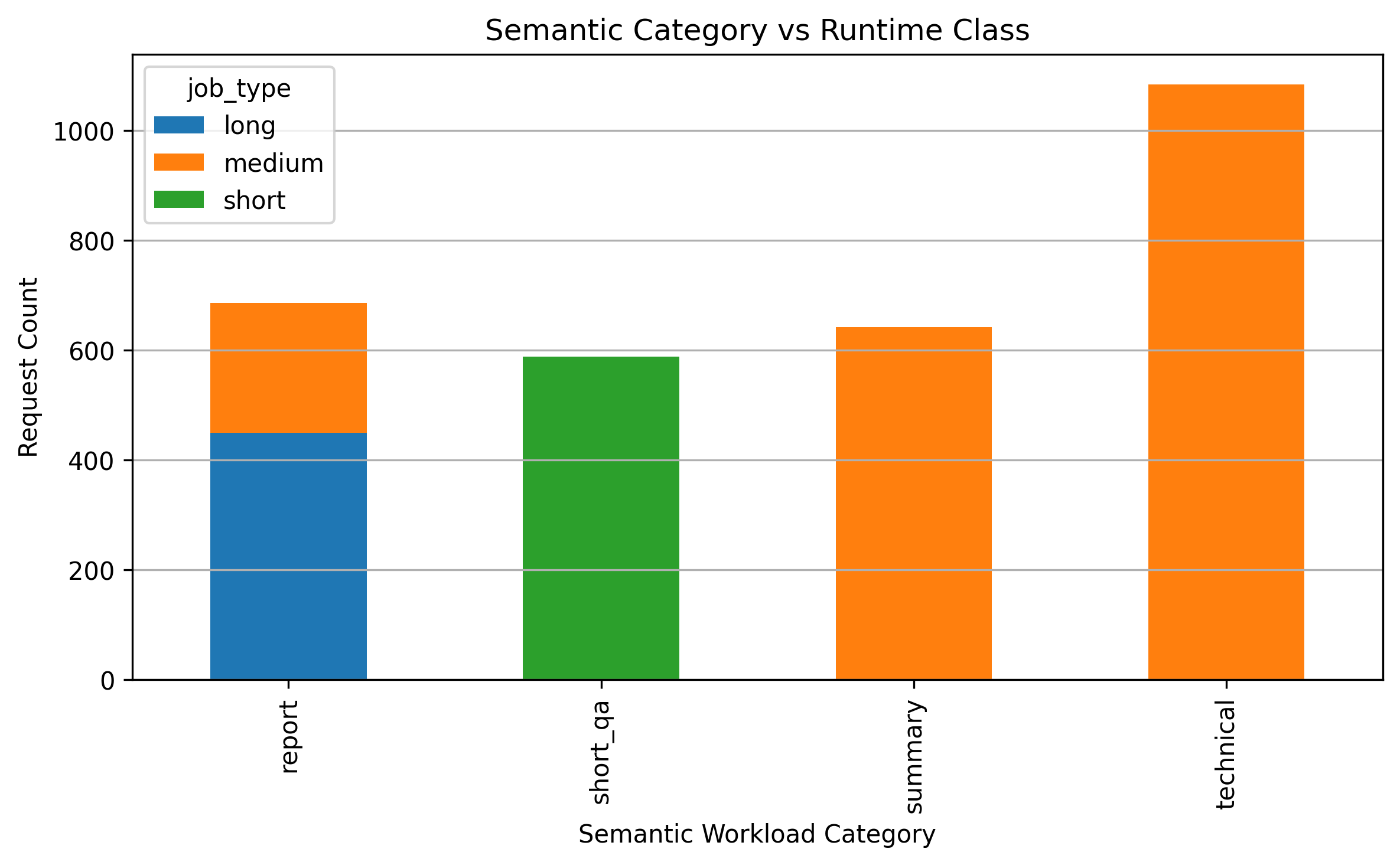}
\caption{SJF}
\end{subfigure}
\hspace{0.04\textwidth}
\begin{subfigure}{0.32\textwidth}
\centering
\includegraphics[width=\linewidth]{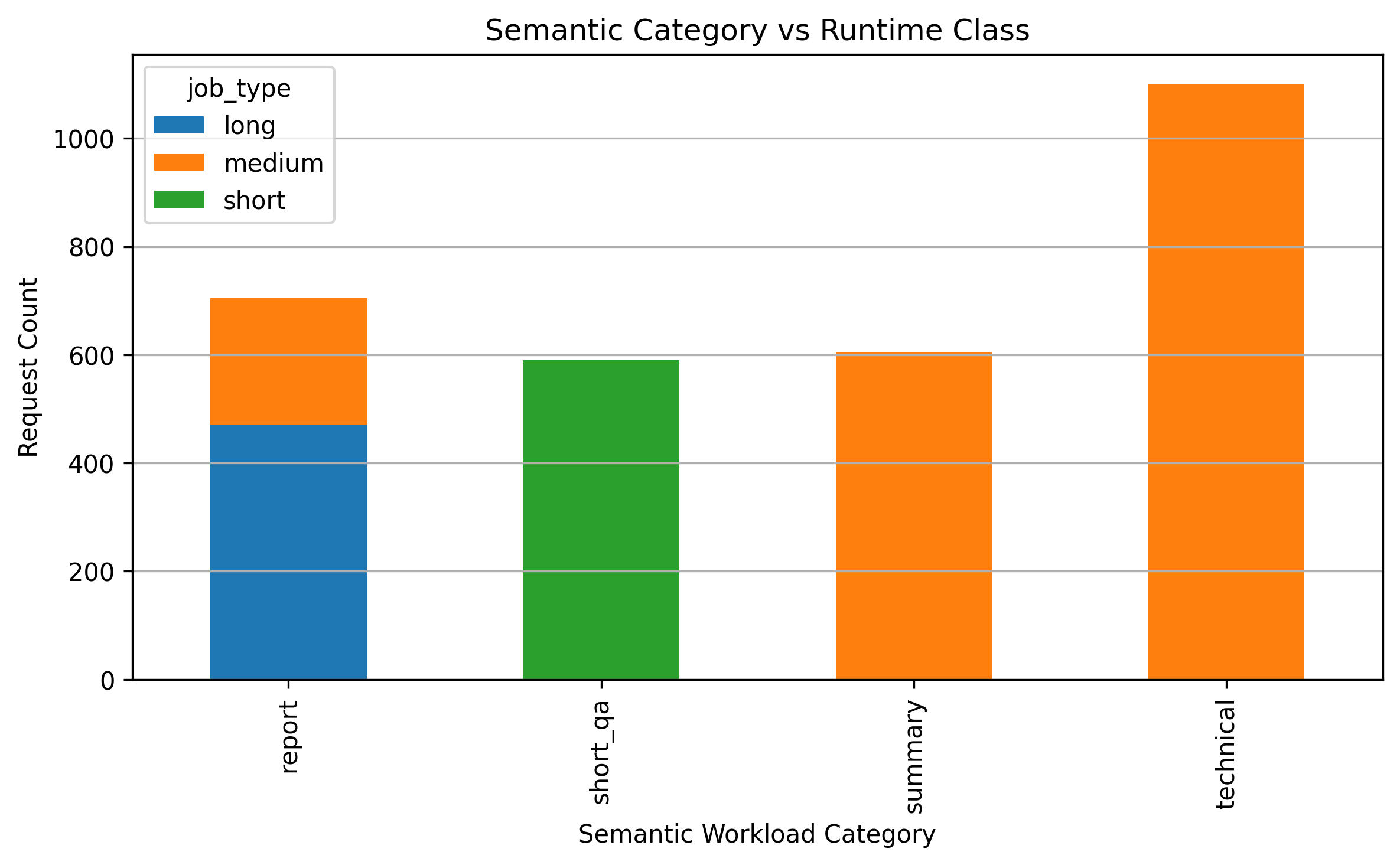}
\caption{Aging Priority}
\end{subfigure}
}

\caption{
Relationship between semantic workload categories and runtime scheduling classes using tokenizer-aware workload characterization. Compared with the whitespace-based workload estimator, tokenizer-aware accounting shifts a substantially larger fraction of report-generation requests into the long-runtime category. This behavior indicates that word-count approximations systematically underestimate the computational cost of certain workload classes. Accurate token-space workload characterization therefore provides a more representative estimate of runtime execution cost and improves admission-time scheduling decisions.
}
\label{fig:semantic_vs_runtime_classification_tokenizer}
\end{figure*}
Experiments were conducted on NVIDIA L4 GPU hardware using vLLM-based inference serving under Ubuntu Linux. Redis was used for distributed queue management and scheduling coordination. Concurrent workload generation was performed using Python-based request injection with configurable concurrency levels to emulate GPU saturation conditions.

The complete DriftSched implementation, experimental automation scripts, runtime metrics pipeline, workload corpus, and scheduling framework are publicly available for reproducibility and future research extensions~\cite{driftschedgithub}.

\subsection{Experimental Configuration}

The experimental configuration consisted of 3000 inference requests, a concurrency level of 50 clients, and three independent experimental runs. Experiments were conducted using a GPU batch size of 32 requests and a batch wait interval of 0.01 seconds before dispatching requests to the GPU inference worker.

To evaluate the impact of workload-estimation fidelity and runtime calibration, two workload-characterization configurations were evaluated. The first utilized a whitespace-delimited proxy (\texttt{split()}) as a controlled fault-injection baseline. The second utilized tokenizer-aware accounting using the model's native tokenizer. For each workload-characterization strategy, experiments were conducted with runtime calibration disabled (BIAS=OFF) and enabled (BIAS=ON).

\subsection{Experimental Hardware and Software Environment}

Experiments were conducted on a bare-metal NVIDIA L4 inference platform to evaluate workload-aware QoS scheduling behavior under concurrent multi-tenant LLM inference workloads. The experimental environment was designed to emulate realistic GPU contention scenarios commonly observed in enterprise AI serving infrastructure. All scheduling policies, workload generation, runtime telemetry collection, and inference execution were performed on the same system to ensure consistent runtime measurements across experiments.

The hardware platform consisted of an NVIDIA L4 inference accelerator~\cite{nvidial4} deployed on an Intel Xeon 6 (Granite Rapids) server platform with DDR5 memory and NVMe-based storage. GPU telemetry including utilization, memory consumption, power draw, temperature, and clock frequencies was collected using NVIDIA-SMI during workload execution. The complete experimental platform configuration is summarized in Table~\ref{tab:experimental_platform}. Hardware specifications are presented in Table~\ref{tab:hardware_specs}, while the software environment configuration is shown in Table~\ref{tab:software_specs}.

\subsection{Admission-Time Workload Characterization Overhead}

Accurate workload characterization is an important component of admission-time scheduling because workload estimates directly influence queue placement, runtime classification, and scheduler decisions. While tokenizer-aware accounting provides a more representative estimate of execution cost than lightweight whitespace-based approximations, it also introduces additional preprocessing overhead at the ingestion layer.

To quantify this tradeoff, we measured the average admission-time workload characterization cost across approximately 295 prompts from each workload category. Measurements compare whitespace-based estimation using \texttt{split()} against tokenizer-aware accounting using the native tokenizer of the Qwen1.5-1.8B-Chat model. Average processing time was computed over repeated executions and reported in milliseconds.

\begin{table}[t]
\centering
\caption{Admission-Time Workload Characterization Overhead}
\label{tab:tokenizer_overhead}
\begin{tabular}{lcccc}
\hline
Category & Count &
\texttt{split()} (ms) &
Tokenizer (ms) &
Overhead \\
\hline
short\_qa  & 295 & 0.00046 & 0.02413 & 52.46$\times$ \\
summary    & 295 & 0.00047 & 0.02436 & 52.22$\times$ \\
technical  & 295 & 0.00049 & 0.02518 & 51.40$\times$ \\
report     & 295 & 0.00059 & 0.03336 & 56.23$\times$ \\
\hline
Average & 295 & 0.00050 & 0.02676 & 53.08$\times$ \\
\hline
\end{tabular}
\end{table}

The results show that tokenizer-aware accounting requires approximately 53$\times$ more CPU processing than whitespace-based estimation. However, the absolute overhead remains extremely small, averaging only 0.027\,ms per request. Compared with the multi-second queue waiting times and inference latencies observed throughout the experiments, this admission-time cost is negligible.

These findings indicate that tokenizer-aware accounting can substantially improve workload-characterization fidelity while imposing only a minimal admission-time processing penalty. Given the negligible absolute overhead, the primary tradeoff between the two approaches is workload-estimation accuracy rather than admission-time processing cost. Whitespace-based estimation offers a lightweight approximation that may require adaptive correction, whereas tokenizer-aware accounting provides higher-fidelity workload estimates at the expense of a modest increase in admission-time CPU processing.

\section{Results and Analysis}

\subsection{Experimental Overview}

Experiments were conducted on an NVIDIA L4 GPU using the vLLM inference runtime. A total of 3000 inference requests were generated across Premium, Standard, and Batch tenants. Workloads consisted of short question-answering, summarization, technical explanation, and long-form report generation tasks. Five scheduling policies were evaluated: FIFO, Priority Scheduling, Shortest Job First (SJF), Weighted Scheduling, and Aging Priority Scheduling.

Each experiment consisted of a 1000-request calibration phase followed by a 2000-request stress evaluation phase. During the calibration phase, runtime execution observations were collected and workload-specific calibration factors were updated when runtime calibration was enabled. The stress phase evaluated scheduler behavior under sustained GPU contention while continuing to collect runtime feedback throughout execution.

\subsection{Semantic Workloads versus Runtime Scheduling Classes}

Figures~\ref{fig:semantic_vs_runtime_classification} and \ref{fig:semantic_vs_runtime_classification_tokenizer} illustrate the relationship between semantic workload categories and runtime scheduling classes under whitespace-based workload estimation (\texttt{split()}) and tokenizer-aware workload characterization, respectively.

The results show that semantic workload categories do not always correspond directly to runtime execution cost. Across all scheduling policies, \texttt{short\_qa} workloads are consistently classified as short jobs, while summary workloads are predominantly medium. Report and technical workloads exhibit greater variability, indicating that semantically similar requests may incur substantially different execution costs.

Compared with whitespace-based workload estimation, tokenizer-aware accounting shifts a larger fraction of report workloads into the long-runtime category across all evaluated scheduling policies. This behavior indicates that word-count approximations systematically underestimate token-space execution cost for certain workload classes.

Similar trends are observed across FIFO, Priority, Weighted, SJF, and Aging Priority scheduling, indicating that the differences are primarily caused by workload-characterization fidelity rather than scheduler-specific queue dynamics.

\begin{figure*}[t]
\centering

\begin{subfigure}{0.32\textwidth}
\centering
\includegraphics[width=\linewidth]{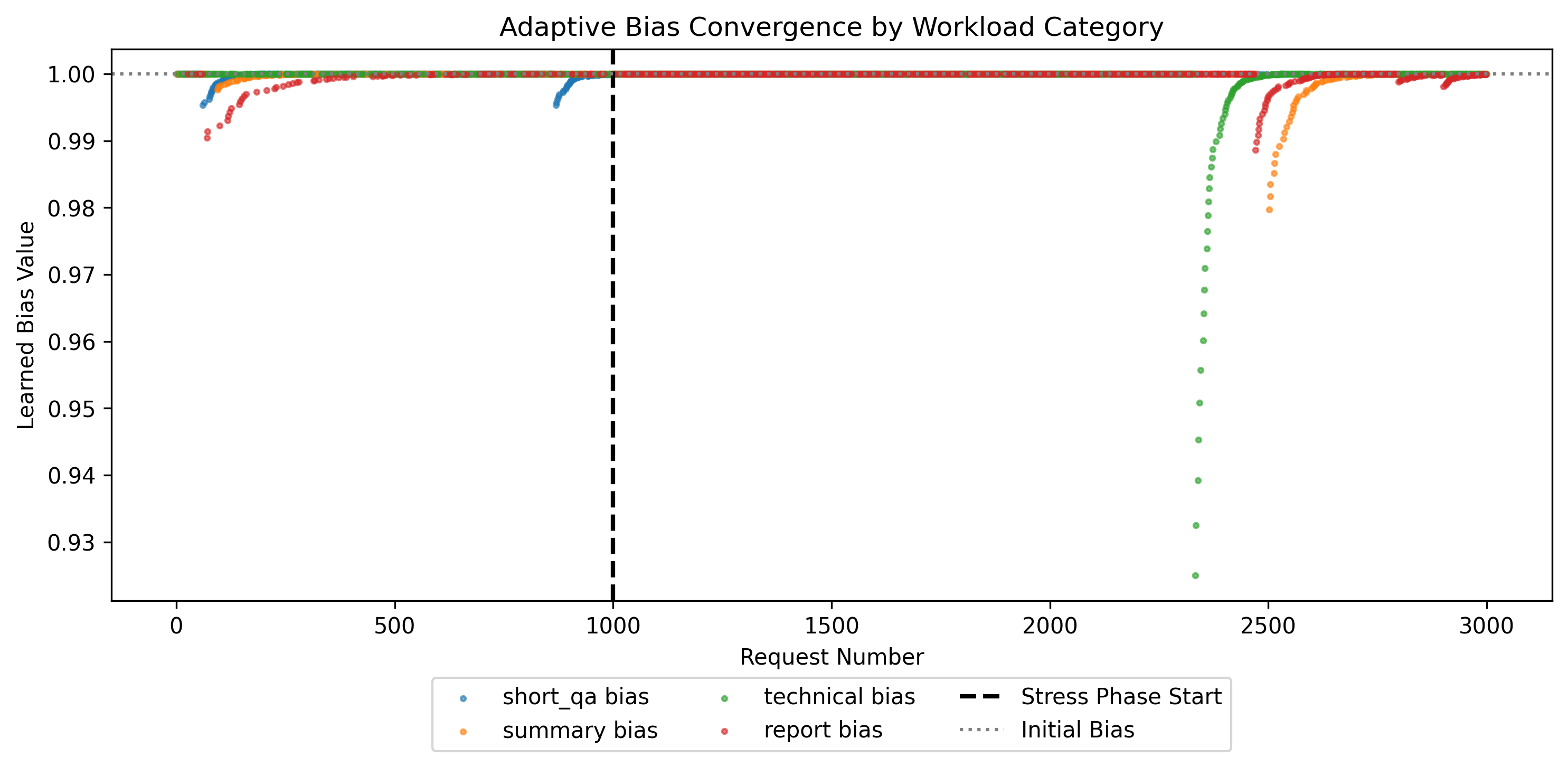}
\caption{FIFO}
\end{subfigure}
\hfill
\begin{subfigure}{0.32\textwidth}
\centering
\includegraphics[width=\linewidth]{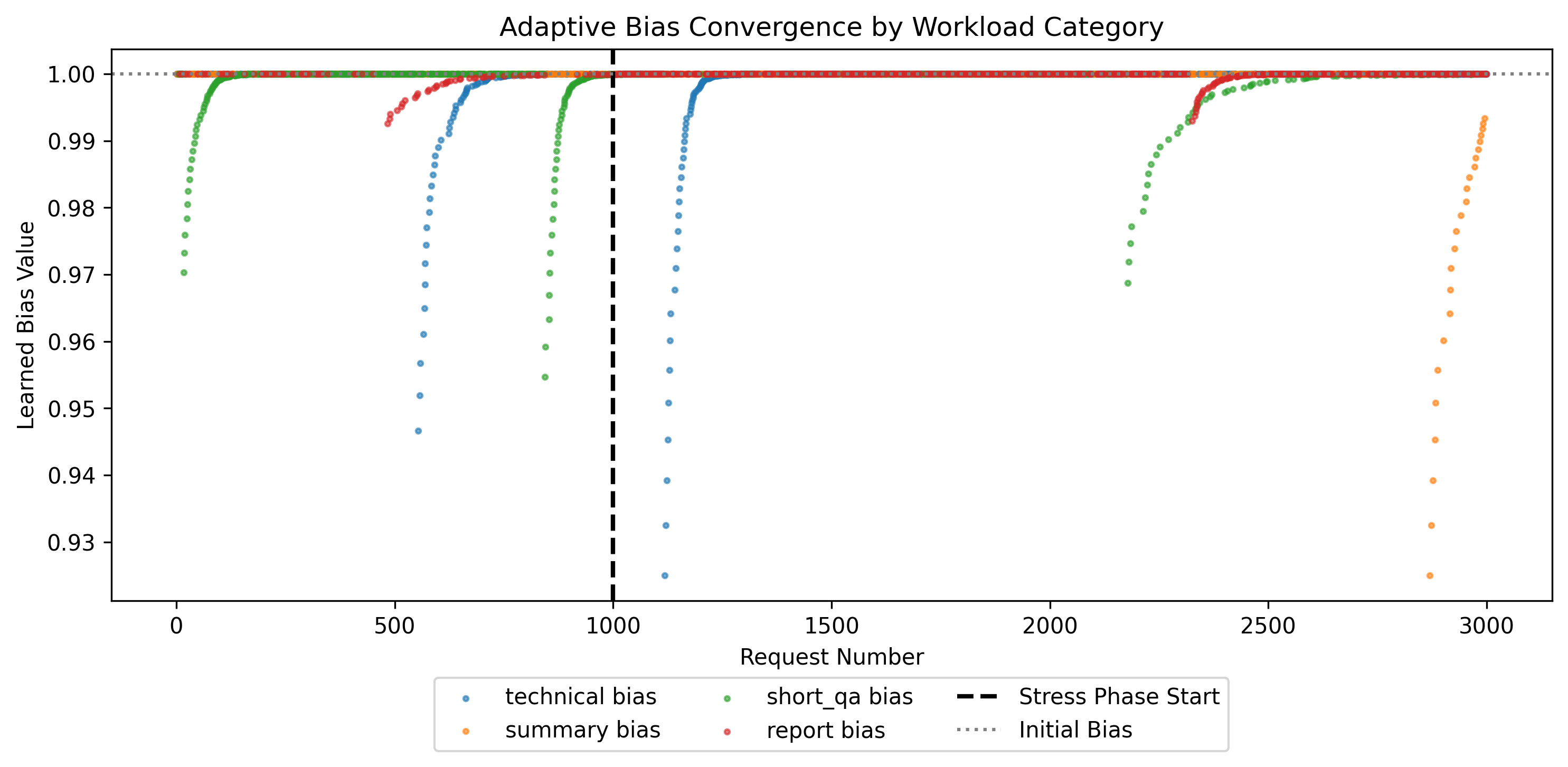}
\caption{Priority }
\end{subfigure}
\hfill
\begin{subfigure}{0.32\textwidth}
\centering
\includegraphics[width=\linewidth]{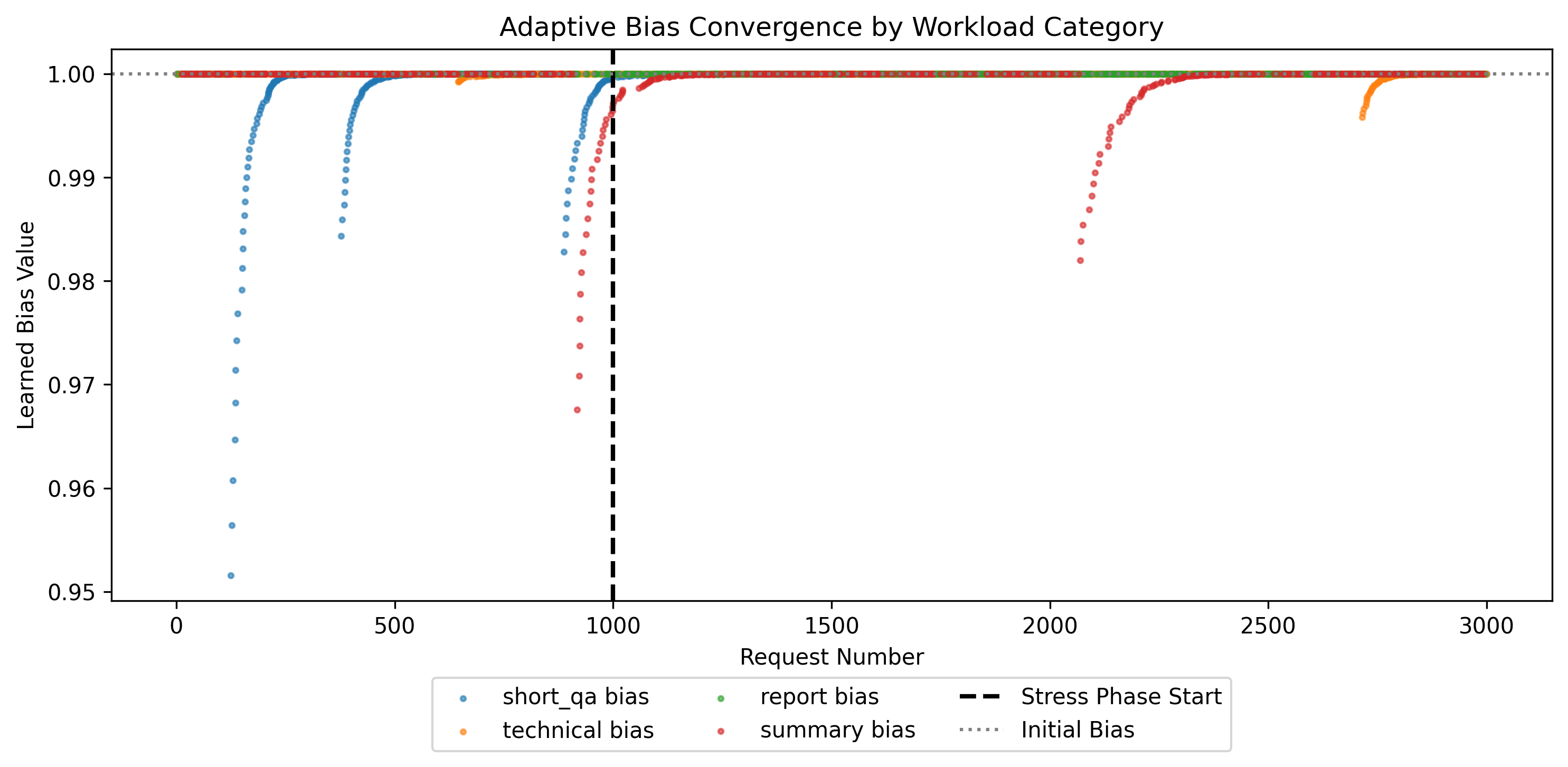}
\caption{Weighted}
\end{subfigure}

\vspace{0.5em}

\makebox[\textwidth][c]{%
\begin{subfigure}{0.38\textwidth}
\centering
\includegraphics[width=\linewidth]{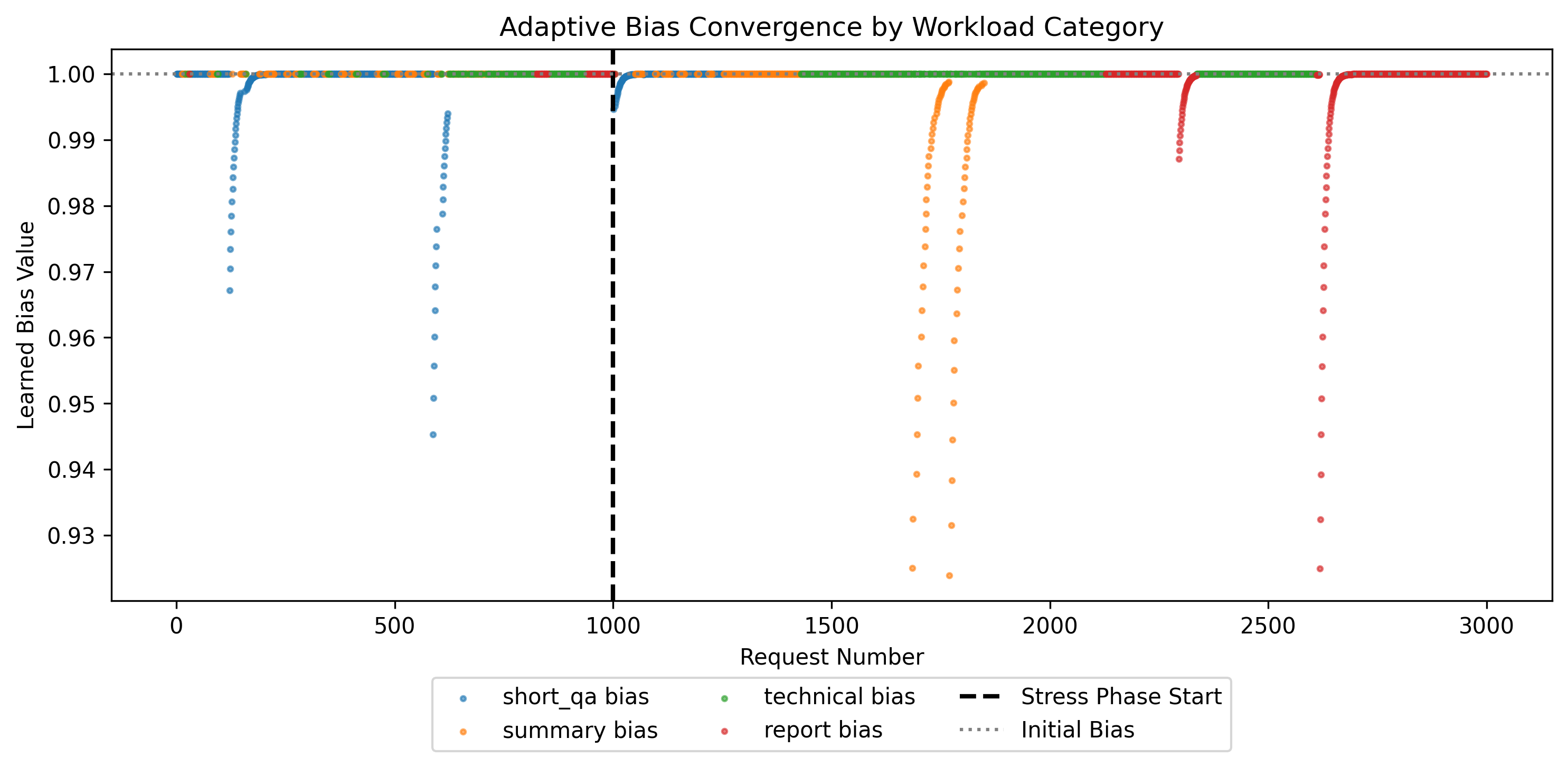}
\caption{SJF}
\end{subfigure}
\hspace{0.04\textwidth}
\begin{subfigure}{0.38\textwidth}
\centering
\includegraphics[width=\linewidth]{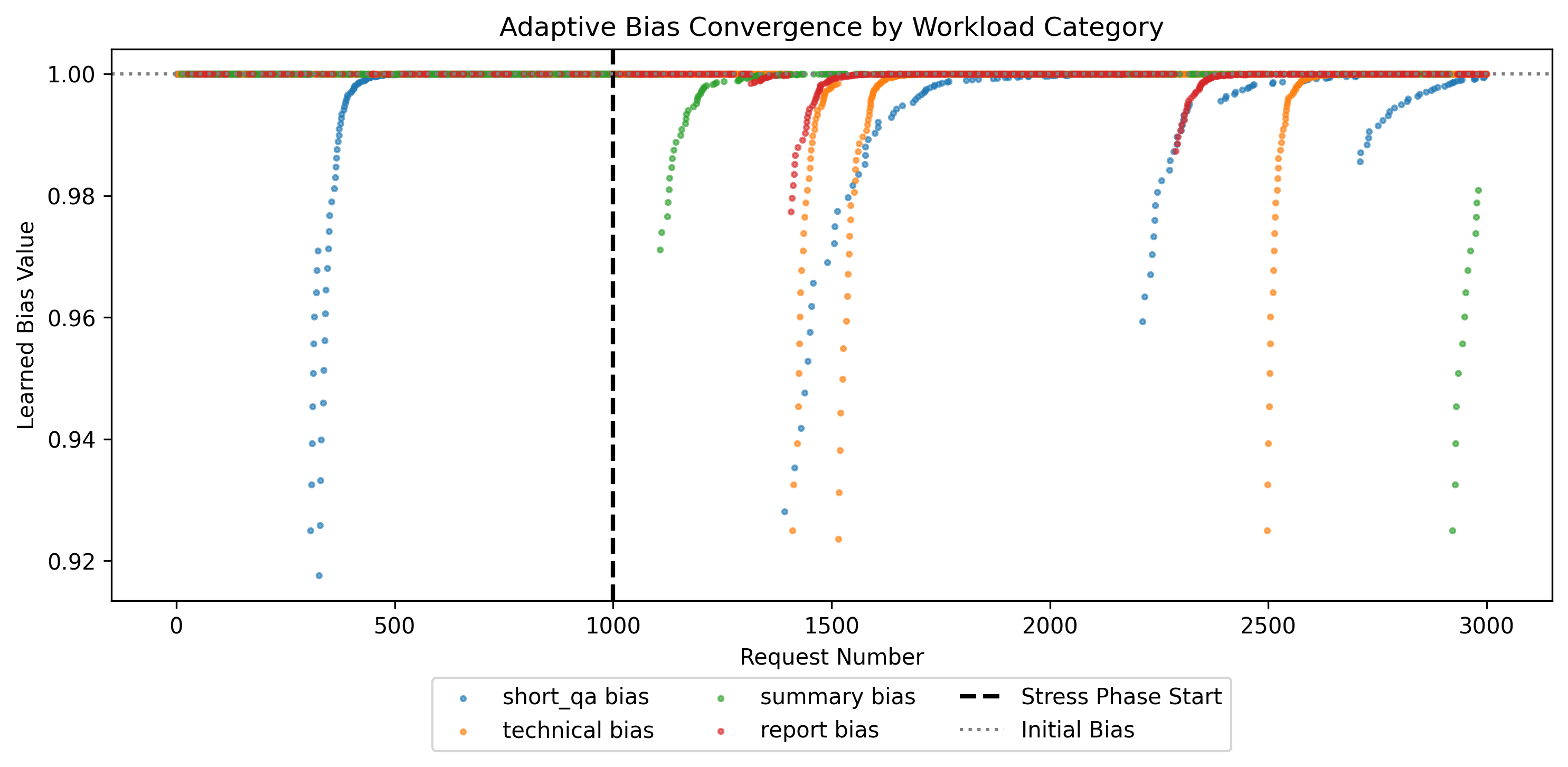}
\caption{Aging Priority}
\end{subfigure}
}

\caption{
Adaptive bias convergence under tokenizer-aware BIAS=ON workload characterization for FIFO, Priority, Weighted, SJF, and Aging Priority scheduling. Bias values remain close to 1.0 throughout execution, indicating that tokenizer-aware workload characterization largely eliminates systematic estimation error and substantially reduces the need for adaptive runtime correction.
}
\label{fig:tokenizer_bias_convergence}
\end{figure*}

\subsection{Runtime Calibration Behavior}

Figure~\ref{fig:bias_convergence_all_schedulers} illustrates the evolution of workload-specific calibration factors during execution.

The results reveal distinct calibration behavior across workload-characterization strategies. For the whitespace-proxy configuration, calibration factors rapidly diverge from the default value of 1.0 and converge toward workload-specific operating regions, indicating the presence of systematic workload-estimation error introduced by word-based approximations. The EMA-based feedback mechanism successfully compensates for these inaccuracies over time by incorporating runtime observations into future workload estimates.

\subsection{Tokenizer-Aware Workload Characterization}

Figure~\ref{fig:tokenizer_bias_convergence} illustrates workload-specific calibration behavior when workload size is estimated using the model's native tokenizer rather than the whitespace-delimited proxy used in earlier experiments.

Across all evaluated scheduling policies, workload-specific calibration factors remain close to unity throughout execution. Minor fluctuations during the calibration phase quickly stabilize, indicating that tokenizer-aware accounting already provides accurate admission-time estimates. In contrast to the whitespace-based configuration, only minimal runtime correction is required, suggesting that most previously observed drift originates from workload-characterization error rather than intrinsic variability in model generation behavior.

\subsection{Impact of Workload Characterization and Runtime Calibration}

To isolate the contributions of workload-characterization fidelity and adaptive feedback, four configurations were evaluated: whitespace-based workload characterization with and without EMA-based calibration, and tokenizer-aware workload characterization with and without adaptive calibration.

\begin{table}[t]
\centering
\caption{Token Estimation Error Across Workload Characterization Modes}
\label{tab:four_mode_token_error}
\begin{tabular}{lccc}
\hline
Configuration & MAE & RMSE & Mean Bias \\
\hline
\texttt{split()} + BIAS=OFF & 70.07 & 97.30 & 0.821 \\
\texttt{split()} + BIAS=ON  & 42.87 & 57.92 & 0.999 \\
Tokenizer + BIAS=OFF & 47.38 & 61.63 & 1.002 \\
Tokenizer + BIAS=ON & 47.06 & 61.32 & 1.003 \\
\hline
\end{tabular}
\end{table}

Table~\ref{tab:four_mode_token_error} summarizes the impact of workload characterization and adaptive runtime calibration on token-estimation accuracy. Under whitespace-based workload characterization, enabling EMA-based calibration reduces MAE from 70.07 to 42.87 tokens and RMSE from 97.30 to 57.92 tokens, while shifting the mean bias from 0.821 toward unity. These results demonstrate that adaptive feedback effectively compensates for systematic estimation errors introduced by coarse workload characterization.

In contrast, tokenizer-aware accounting exhibits nearly identical estimation accuracy with and without adaptive calibration. The MAE and RMSE differ by less than one token, and mean bias remains approximately one in both configurations. This indicates that tokenizer-aware workload characterization already provides stable admission-time estimates and requires little additional runtime correction.

Overall, these results suggest that adaptive calibration is most beneficial when workload estimation is coarse, whereas tokenizer-aware accounting largely eliminates systematic estimation drift. Once workload estimates become sufficiently accurate, scheduler selection becomes the dominant factor influencing latency and tenant-level QoS behavior.
\begin{table}[t]
\centering
\caption{Summary of Observations Across the Four Configurations}
\label{tab:four_mode_summary}
\begin{tabular}{lccc}
\hline
Configuration & Bias & EMA Benefit & Observation \\
\hline
split()+OFF &
0.821 &
None &
Largest drift \\

split()+ON &
$\approx1$ &
High &
EMA corrects errors \\

Tokenizer+OFF &
$\approx1$ &
Low &
Stable estimates \\

Tokenizer+ON &
$\approx1$ &
Minimal &
Nearly identical \\
\hline
\end{tabular}
\end{table}

Table~\ref{tab:four_mode_summary} summarizes the qualitative observations obtained from the four workload-characterization configurations. The results indicate that EMA-based calibration provides the greatest benefit when workload estimation relies on coarse whitespace-based approximations. In contrast, tokenizer-aware accounting produces stable admission-time estimates with or without runtime calibration, resulting in nearly identical behavior between BIAS=OFF and BIAS=ON configurations.

\subsection{Tenant Queue Dynamics Under Contention}

\begin{figure*}[t]
\centering

\begin{subfigure}{0.32\textwidth}
\centering
\includegraphics[width=\linewidth]{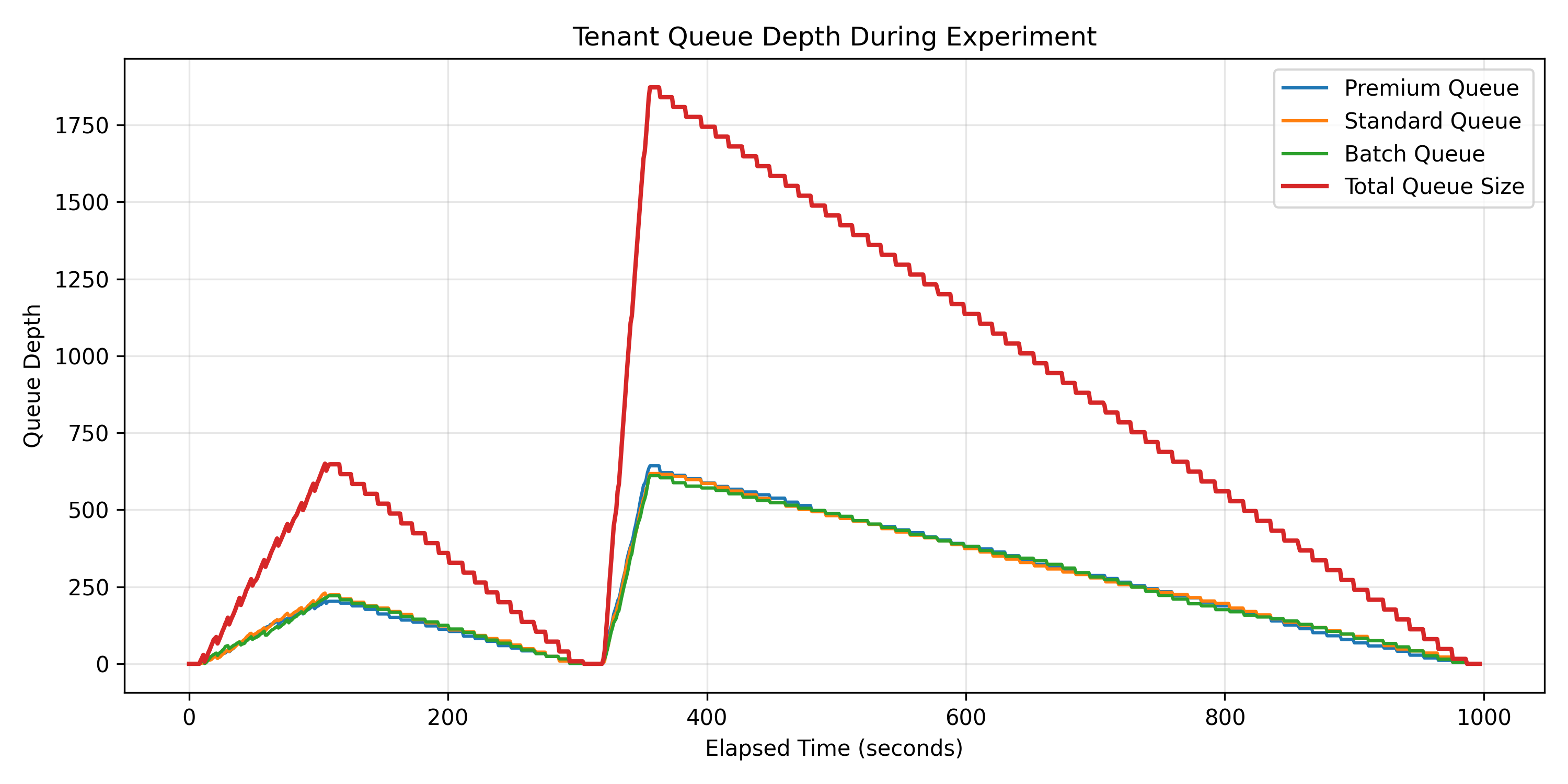}
\caption{FIFO}
\end{subfigure}
\hfill
\begin{subfigure}{0.32\textwidth}
\centering
\includegraphics[width=\linewidth]{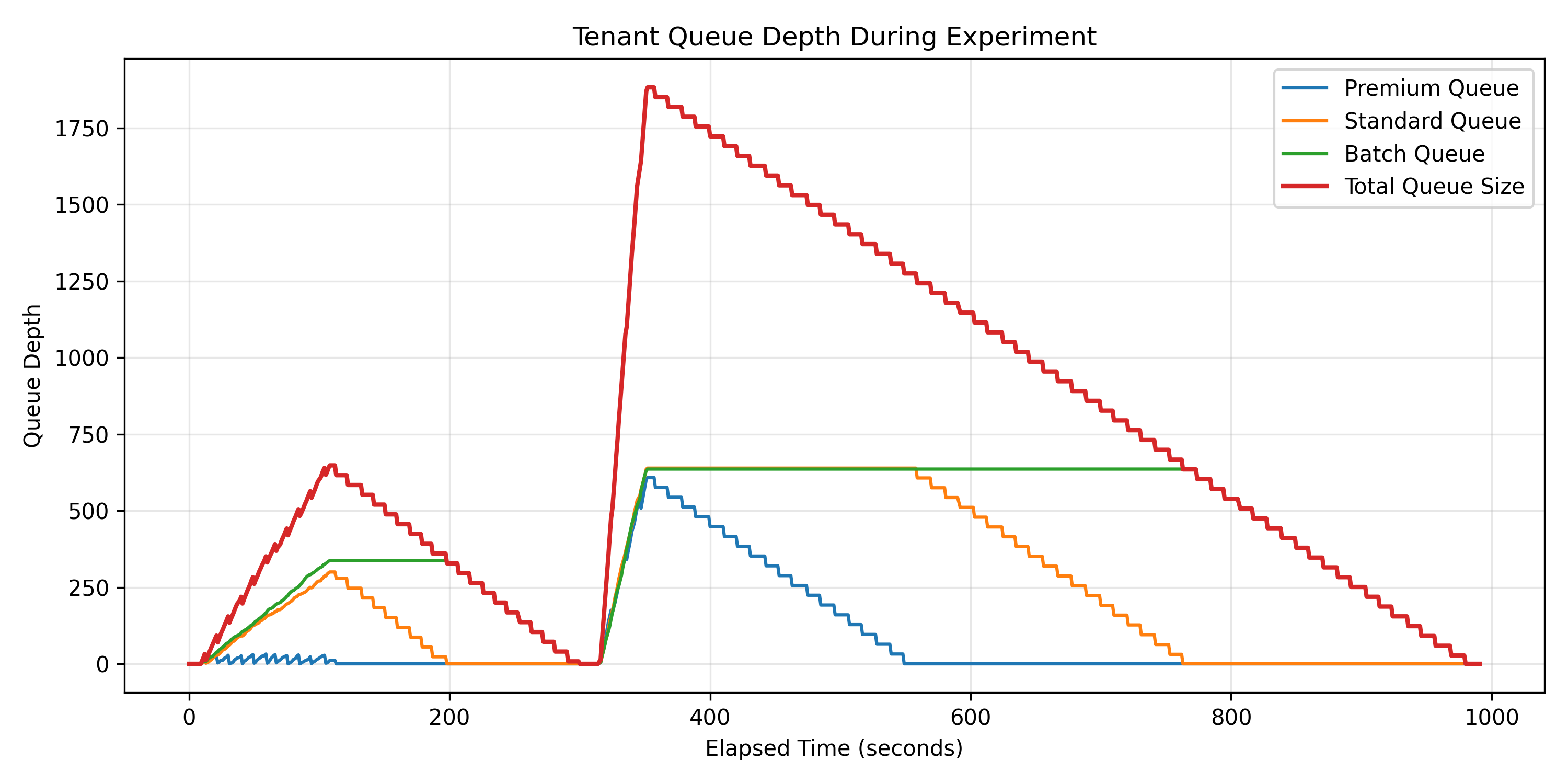}
\caption{Priority }
\end{subfigure}
\hfill
\begin{subfigure}{0.32\textwidth}
\centering
\includegraphics[width=\linewidth]{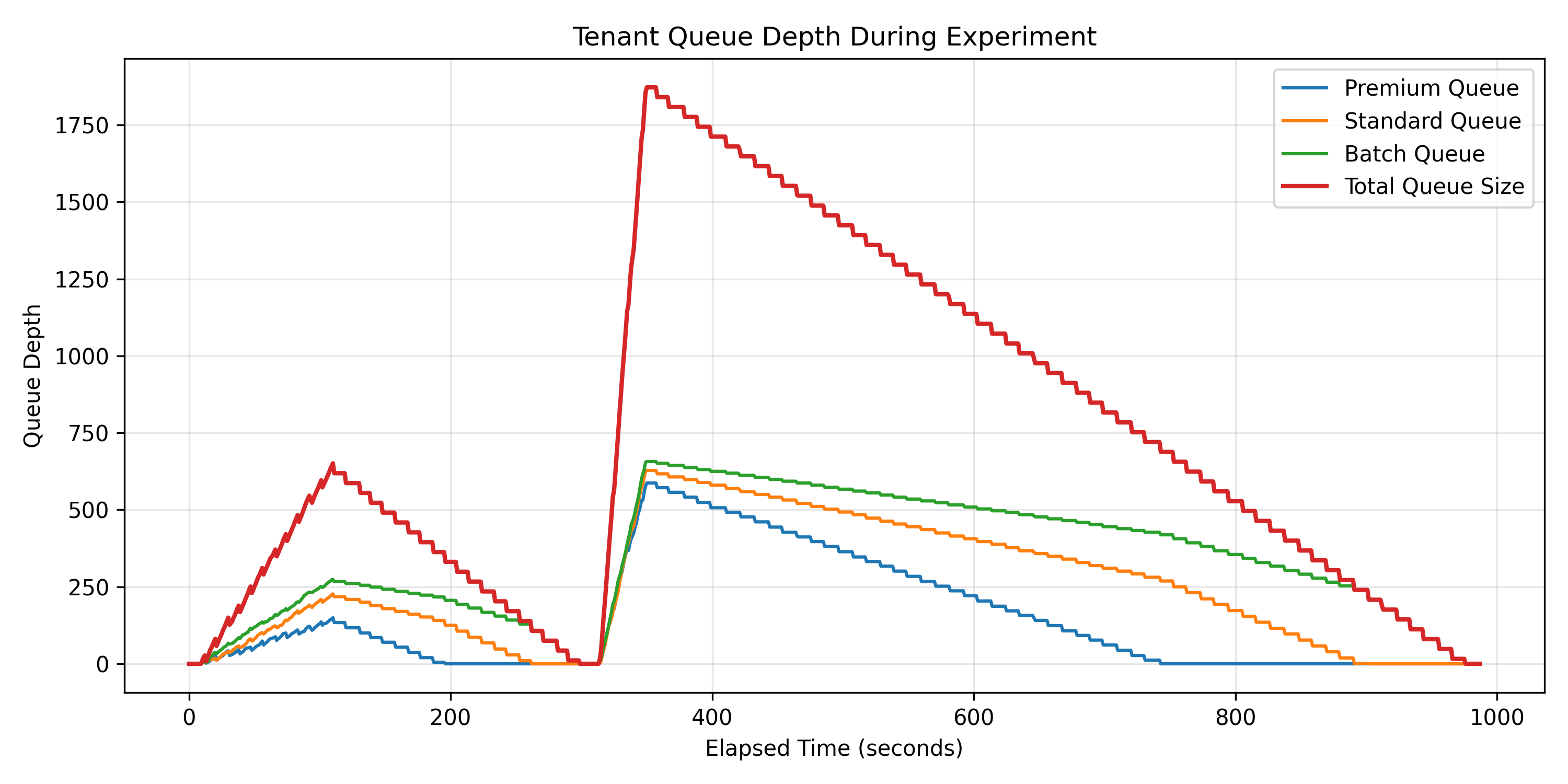}
\caption{Weighted}
\end{subfigure}
\vspace{0.5em}
\makebox[\textwidth][c]{%
\begin{subfigure}{0.32\textwidth}
\centering
\includegraphics[width=\linewidth]{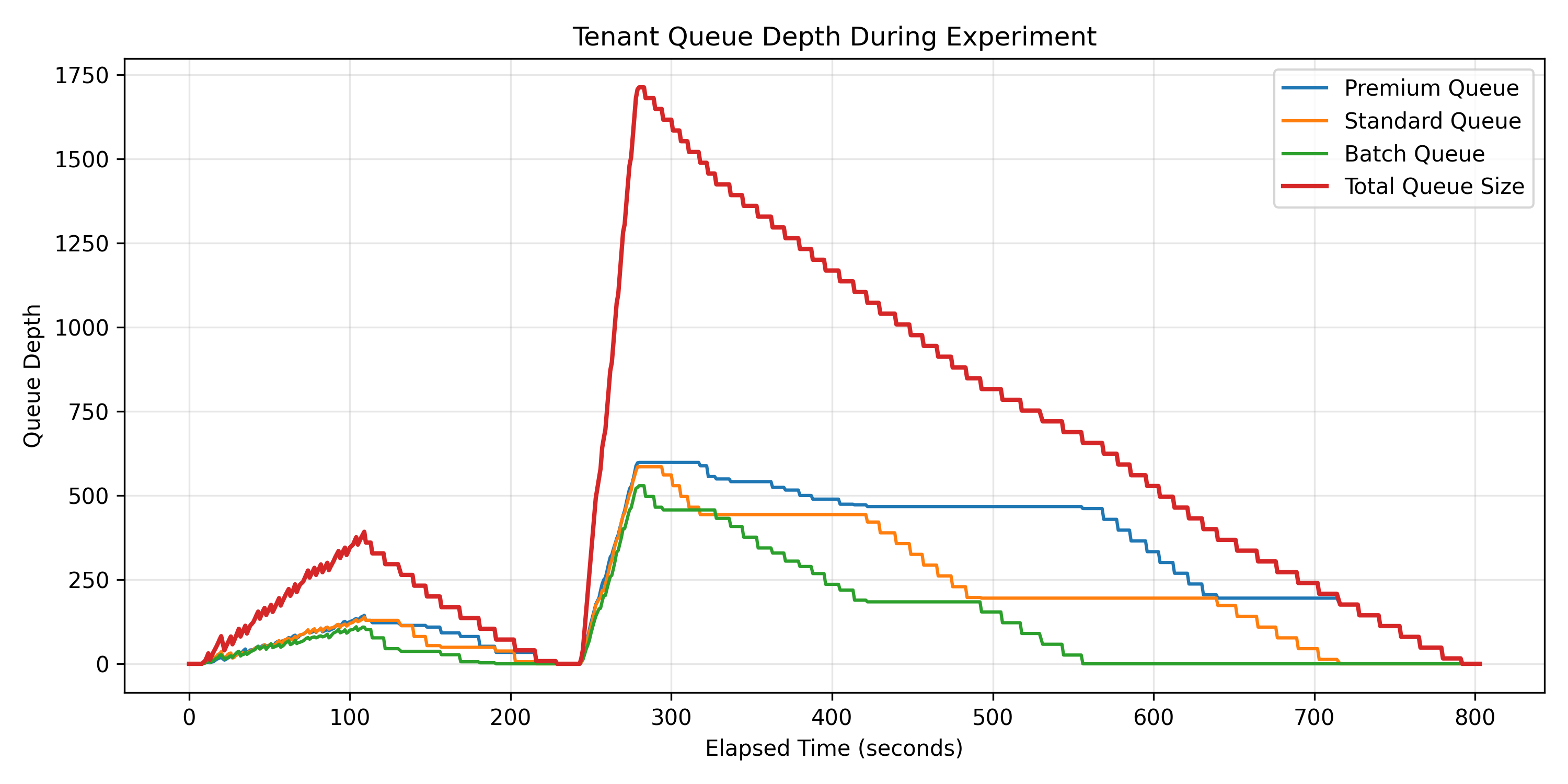}
\caption{SJF}
\end{subfigure}
\hspace{0.04\textwidth}
\begin{subfigure}{0.38\textwidth}
\centering
\includegraphics[width=\linewidth]{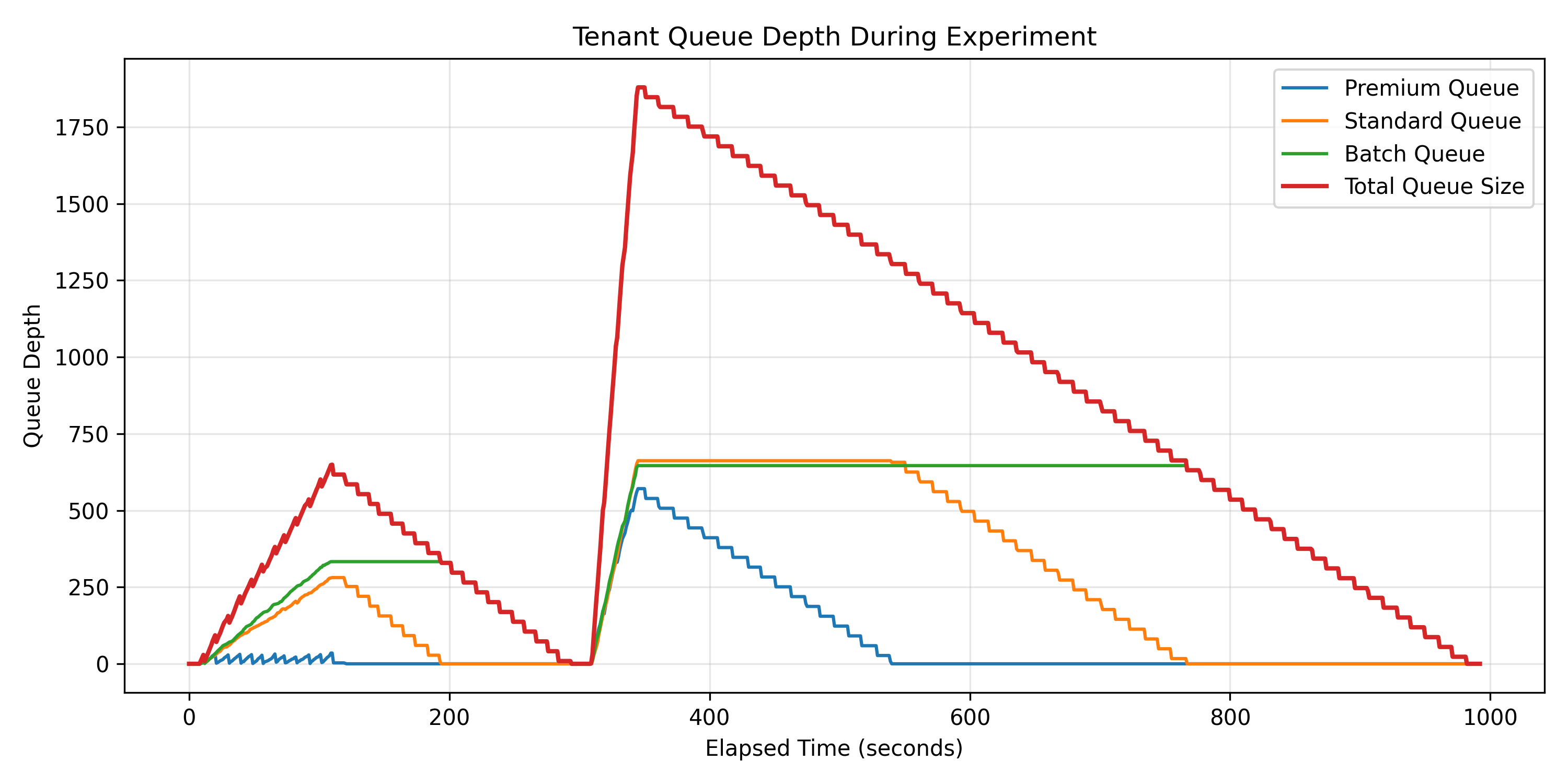}
\caption{Aging Priority}
\end{subfigure}
}

\caption{
Tenant queue depth evolution for FIFO, Priority, Weighted, SJF, and Aging Priority scheduling under sustained multi-tenant GPU contention. FIFO provides uniform queue draining, Priority and SJF favor higher-priority or shorter workloads, while Weighted and Aging Priority provide more balanced queue progression across tenants.
}
\label{fig:queue_depth_all_schedulers}
\end{figure*}

Figure~\ref{fig:queue_depth_all_schedulers} illustrates queue depth evolution for Premium, Standard, and Batch tenants across all evaluated scheduling policies. The red curve represents total queue depth, while individual tenant queues are shown separately.

Queue depth evolution reflects the transition from the 1000-request calibration phase to the subsequent 2000-request stress phase. Under sustained GPU contention, queue buildup occurs across all tenant classes, providing a realistic environment for evaluating scheduler behavior.

The results reveal substantial differences in queue management behavior under sustained GPU contention. FIFO scheduling exhibits nearly identical queue growth and drain patterns across all tenant classes, providing fairness but no service differentiation. In contrast, Priority Scheduling aggressively favors Premium workloads, causing high-priority queues to drain rapidly while lower-priority Batch workloads remain queued for longer periods.

Weighted Scheduling provides a more balanced allocation strategy. Although Premium tenants continue to receive preferential treatment, all tenant queues make continuous progress, reducing starvation risk while preserving QoS differentiation. Aging Priority Scheduling further improves fairness by gradually increasing the priority of long-waiting requests, allowing Batch workloads to eventually execute even under sustained high-priority traffic.

SJF demonstrates the most aggressive queue reduction behavior due to its preference for shorter estimated workloads. The total queue depth decreases more rapidly than in FIFO and Priority Scheduling, indicating improved queue efficiency. Because SJF relies directly on estimated workload size, its effectiveness is strongly influenced by workload-characterization fidelity.

Across all schedulers, the stress phase produces a substantial increase in queue depth as request arrival rates exceed instantaneous GPU processing capacity. The resulting queue buildup provides a realistic evaluation environment for studying scheduler behavior under multi-tenant GPU saturation conditions.

Similar queue-draining patterns were observed under tokenizer-aware workload characterization and are omitted for brevity. While accurate token-space workload estimation improves admission-time workload classification, the overall queue evolution remains primarily governed by the selected scheduling policy. Despite differences in workload-characterization fidelity, FIFO, Priority, Weighted, SJF, and Aging Priority exhibit qualitatively similar queue draining behavior under both configurations.

\subsection{Tail Latency Analysis}
\begin{figure}[t]
\centering
\includegraphics[width=\linewidth]{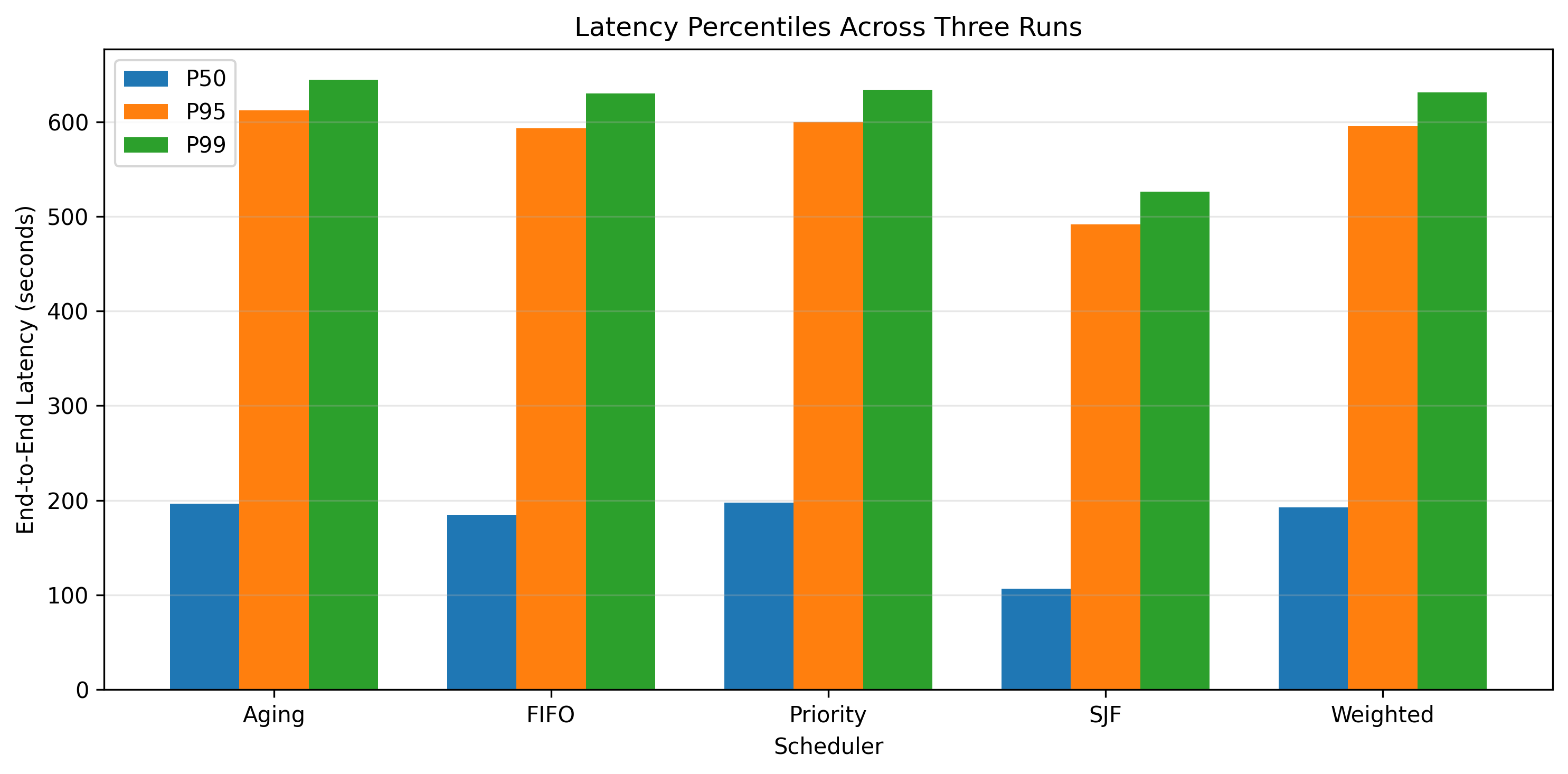}
\caption{ End-to-end latency comparison across scheduling policies using whitespace-based workload characterization (\texttt{split()}). Reported values represent the average P50, P95, and P99 latency across three experimental runs. }
\label{fig:tail_latency_p50_p95_p99}
\end{figure}

Figure~\ref{fig:tail_latency_p50_p95_p99} compares end-to-end latency percentiles across all evaluated scheduling policies using the whitespace-based workload characterization baseline (\texttt{split()}). Tail-latency behavior is particularly important in large-scale serving systems because a small number of delayed requests can significantly impact overall user experience. Dean and Barroso demonstrated that tail latency amplification becomes increasingly severe as system scale and concurrency increase~\cite{tailatscale}. Results are reported using the median (P50), tail (P95), and extreme-tail (P99) latency metrics.

FIFO, Priority, Weighted, and Aging Priority scheduling exhibit broadly similar latency behavior, with P95 latency values near 600 seconds and P99 latency values exceeding 630 seconds. Although Priority and Weighted Scheduling provide service differentiation across tenant classes, their impact on aggregate tail latency remains limited under sustained GPU contention. These results indicate that tenant-aware scheduling alone is insufficient to substantially reduce overall queue buildup when workload characterization is derived from a coarse word-count approximation.

SJF achieves the lowest latency across all reported percentiles, reducing median latency to approximately 107 seconds while lowering P95 and P99 latency to approximately 491 and 526 seconds, respectively. Even under imperfect workload characterization, SJF successfully prioritizes smaller workloads and reduces queue residence time for latency-sensitive requests. This demonstrates that workload-aware scheduling provides substantial latency benefits despite admission-time estimation errors.

\begin{figure*}[t]
\centering

\begin{subfigure}{0.48\textwidth}
\centering
\includegraphics[width=\linewidth]{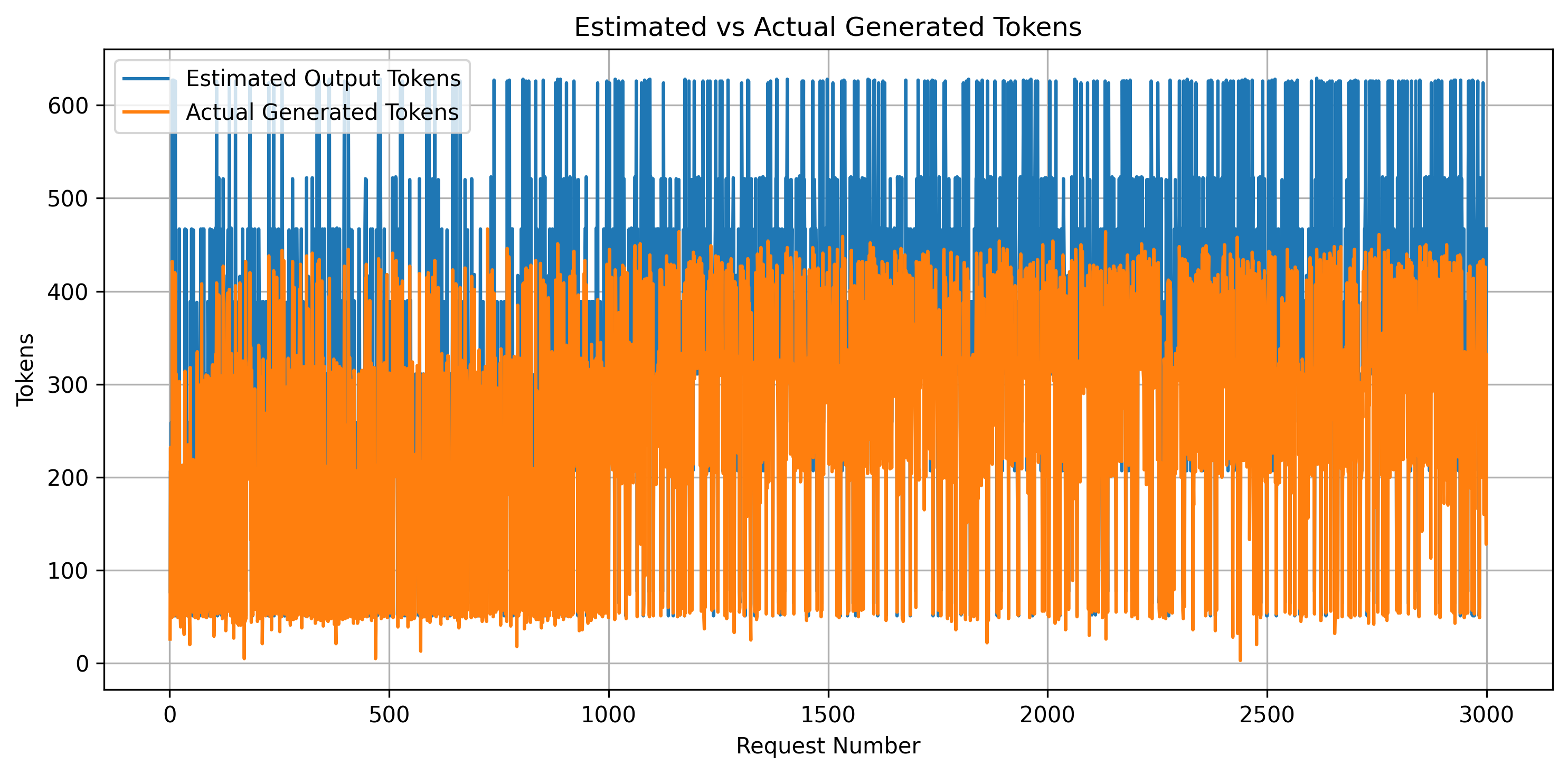}
\caption{FIFO without adaptive runtime token drift compensation}
\label{fig:fifo_no_bias_tokens}
\end{subfigure}
\hfill
\begin{subfigure}{0.48\textwidth}
\centering
\includegraphics[width=\linewidth]{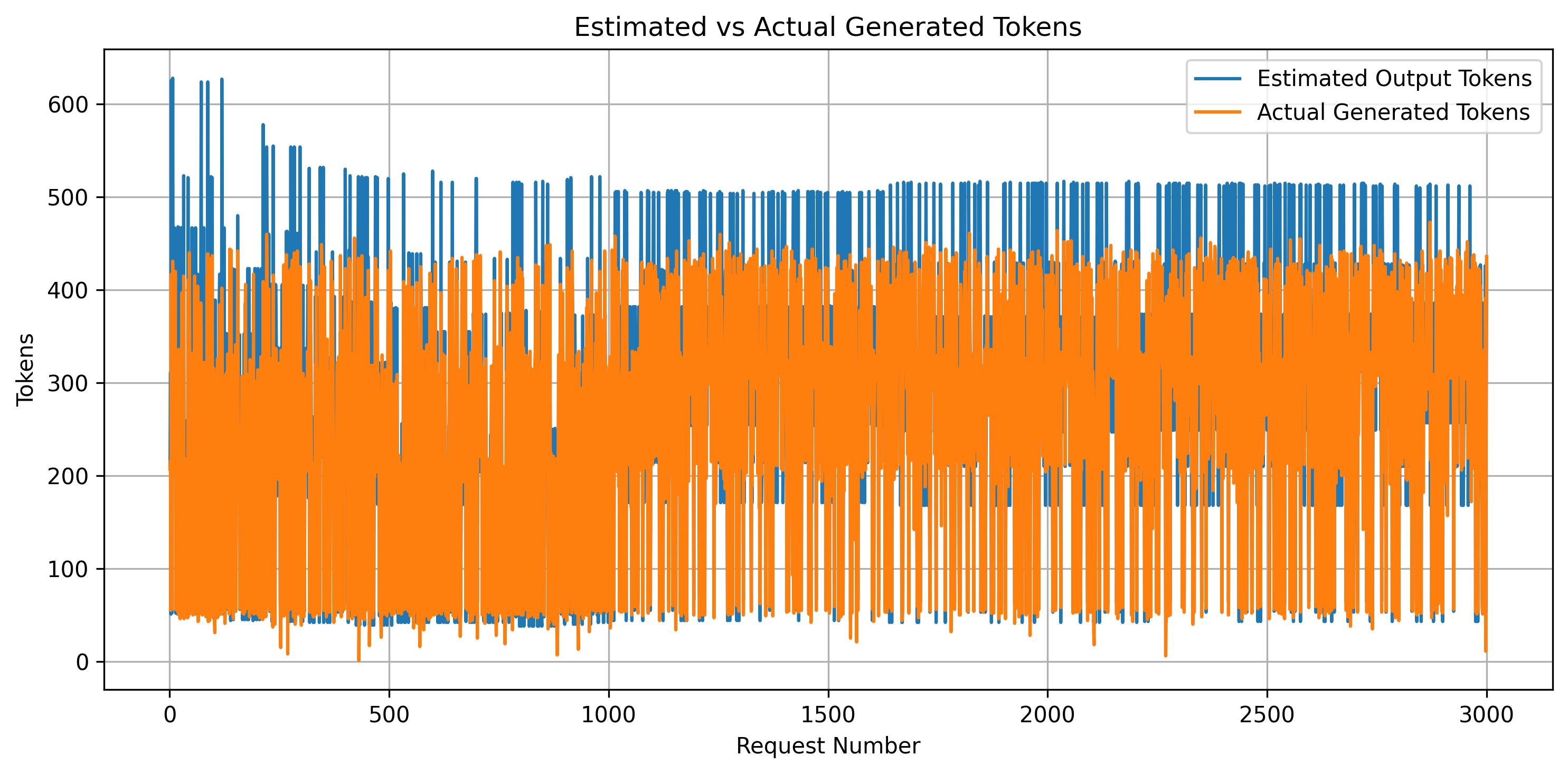}
\caption{FIFO with adaptive runtime token drift compensation}
\label{fig:fifo_bias_tokens}
\end{subfigure}

\caption{
Estimated token budgets versus observed output lengths under FIFO scheduling using whitespace-based workload characterization (\texttt{split()}) with BIAS=OFF and BIAS=ON. Under BIAS=OFF, workload estimates remain based on static word-count assumptions and exhibit larger deviations from observed generation behavior. Under BIAS=ON, EMA-based bias correction progressively adjusts estimated token budgets toward observed output lengths, reducing estimation error and improving workload-classification accuracy over time.
}
\label{fig:token_drift_compensation}
\end{figure*}

SJF achieves the lowest latency across all percentiles. Compared with FIFO, SJF reduces P50 latency by approximately 42\%, while reducing P95 and P99 latency by approximately 17\% and 16\%, respectively. Although tokenizer-aware characterization improves estimation accuracy, the relative ranking of scheduling policies remains unchanged, indicating that scheduler selection has a larger influence on tail latency than workload-estimation refinement alone.

Across all schedulers, the gap between P50 and P99 latency remains substantial, indicating that queueing effects dominate overall response time under sustained multi-tenant load. These results highlight that tail-latency reduction remains a critical objective for future adaptive GPU inference scheduling systems.

\begin{table}[htbp]
\caption{Tail Latency Comparison Across Scheduling Policies Using Whitespace-Based Workload Characterization (\texttt{split()}) (3-Run Average)}
\label{tab:tail_latency}
\centering
\small
\begin{tabular}{lcccc}
\hline
Scheduler & P95 & $\sigma_{95}$ & P99 & $\sigma_{99}$ \\
\hline
FIFO     & 592.957 & 6.686 & 630.205 & 1.502 \\
Priority & 599.760 & 1.738 & 633.684 & 1.792 \\
Weighted & 595.601 & 2.465 & 631.305 & 2.715 \\
SJF      & 491.480 & 3.995 & 526.363 & 5.028 \\
Aging    & 611.968 & 2.472 & 644.645 & 4.905 \\
\hline
\end{tabular}
\end{table}

Table~\ref{tab:tail_latency} summarizes tail-latency behavior across all evaluated scheduling policies using the whitespace-based workload characterization baseline (\texttt{split()}) with adaptive bias compensation enabled. SJF achieved the lowest P95 and P99 latency across all evaluated scheduling policies, reducing P99 latency to 526.4 seconds compared with approximately 630--645 seconds for the remaining schedulers. Relative to FIFO, SJF reduced P95 latency by approximately 17\% and P99 latency by approximately 16\%, demonstrating the effectiveness of workload-size-aware scheduling even when admission-time workload estimation is derived from a coarse word-count proxy.

This behavior is expected because SJF prioritizes shorter requests, reducing queue residence time for latency-sensitive workloads and improving overall latency efficiency. FIFO, Priority, and Weighted Scheduling produced similar tail-latency behavior, with P99 latency remaining near 630 seconds. Aging Priority exhibited the highest P95 and P99 latency because its fairness-oriented promotion mechanism periodically elevates long-waiting requests, increasing tail-latency variability.

The relatively small standard deviations observed across all schedulers indicate stable behavior across repeated experimental runs. Overall, these results demonstrate that workload-aware scheduling can significantly reduce tail latency under sustained multi-tenant GPU inference workloads, although the resulting improvements may come at the expense of tenant-level QoS guarantees.

These results are particularly noteworthy because they were obtained under the more challenging workload-characterization scenario. Although the whitespace-based estimator introduces systematic workload-estimation errors, the adaptive feedback mechanism is able to partially compensate for these inaccuracies and preserve the relative advantages of workload-aware scheduling. As shown later in the tokenizer-aware evaluation, improving workload-characterization fidelity further reduces workload misclassification and improves admission-time scheduling decisions, but does not substantially alter the overall ranking of the evaluated scheduling policies.
\begin{table}[htbp]
\caption{Aggregated Scheduler Performance Across Three Runs Using Split()-Based Workload Estimation}
\label{tab:scheduler_performance}
\centering
\small
\begin{tabular}{lcccc}
\hline
\textbf{Scheduler} &
\textbf{Avg Queue Wait (s)} &
\textbf{P50 (s)} &
\textbf{P95 (s)} &
\textbf{P99 (s)} \\
\hline
FIFO      & 238.8 & 184.7 & 593.0 & 630.2 \\
Priority  & 239.2 & 197.8 & 599.8 & 633.7 \\
Weighted  & 241.0 & 192.8 & 595.6 & 631.3 \\
Aging     & 245.0 & 196.3 & 612.0 & 644.6 \\
SJF       & 149.5 & 106.9 & 491.5 & 526.4 \\
\hline
\end{tabular}
\end{table}

Table~\ref{tab:scheduler_performance} summarizes the aggregated performance of all evaluated scheduling policies across three independent experimental runs using the whitespace-based workload characterization baseline (\texttt{split()}) with adaptive bias compensation enabled. The results show that scheduling policy selection has a significant impact on queue waiting time and tail-latency behavior under sustained multi-tenant GPU contention.

Among all evaluated schedulers, SJF achieves the best overall performance, reducing average queue waiting time to 149.5 seconds compared with approximately 239--245 seconds for the other scheduling policies. SJF also produces the lowest latency across all reported percentiles, reducing median (P50) latency by approximately 42\% relative to FIFO while lowering P95 and P99 latency by approximately 17\% and 16\%, respectively.

FIFO, Priority, Weighted, and Aging Priority scheduling exhibit broadly similar overall latency characteristics. Although Priority, Weighted, and Aging Priority introduce tenant-aware scheduling behavior, their impact on aggregate system latency remains limited. This observation suggests that tenant-priority assignment primarily redistributes service quality among tenant classes rather than reducing overall queueing delay.

The results further indicate that workload-aware scheduling has a greater influence on end-to-end performance than tenant-priority assignment alone. Because SJF prioritizes requests using estimated workload size, improvements in adaptive workload estimation directly translate into improved queue efficiency and lower latency. These findings highlight the importance of accurate runtime workload characterization for reducing queue buildup and improving overall system responsiveness under GPU saturation conditions.

Although SJF achieves the best aggregate latency and queue-wait performance, Table~\ref{tab:tenant_latency} shows that tenant-aware schedulers such as Priority, Weighted, and Aging Priority Scheduling provide substantially stronger QoS differentiation for Premium tenants. These findings reveal a fundamental tradeoff between global latency optimization and tenant-level service guarantees in multi-tenant GPU inference environments.

It is important to note that these results represent the more challenging workload-characterization scenario in which admission-time workload size is estimated using a whitespace-delimited proxy rather than exact tokenizer accounting. The tokenizer-aware workload characterization improves admission-time workload sizing and further reduces workload misclassification, although the relative ranking of the scheduling policies remains largely unchanged.

\begin{table}[htbp]
\caption{Average Tenant-Level End-to-End Latency and Queue Wait Across Three Experimental Runs Using Whitespace-Based Workload Characterization (\texttt{split()})}
\label{tab:tenant_latency}
\centering
\small
\begin{tabular}{lccc}
\hline
\textbf{Scheduler} & \textbf{Tenant} & \textbf{Latency (s)} & \textbf{Queue Wait (s)} \\
\hline
FIFO     & Premium  & 248.23 & 238.04 \\
FIFO     & Standard & 249.25 & 238.93 \\
FIFO     & Batch    & 252.97 & 242.77 \\
\hline
Priority & Premium  & 77.32  & 67.18 \\
Priority & Standard & 252.80 & 242.63 \\
Priority & Batch    & 426.72 & 416.57 \\
\hline
Weighted & Premium  & 158.45 & 148.25 \\
Weighted & Standard & 255.02 & 244.82 \\
Weighted & Batch    & 333.05 & 322.90 \\
\hline
SJF      & Premium  & 226.60 & 218.10 \\
SJF      & Standard & 157.52 & 149.38 \\
SJF      & Batch    & 94.91  & 87.07 \\
\hline
Aging    & Premium  & 76.39  & 66.26 \\
Aging    & Standard & 256.07 & 245.99 \\
Aging    & Batch    & 433.00 & 422.87 \\
\hline
\end{tabular}
\end{table}

Table~\ref{tab:tenant_latency} summarizes average tenant-level latency and queue waiting behavior across three independent experimental runs using the whitespace-based workload characterization baseline (\texttt{split()}) with adaptive bias compensation enabled. FIFO scheduling produces nearly identical latency and queue wait times across all tenant classes, demonstrating strong fairness but providing no explicit QoS differentiation.

Priority Scheduling exhibits the strongest tenant-level service differentiation. Premium tenant latency is reduced to approximately 77 seconds, while Batch tenant latency increases to approximately 427 seconds. This behavior demonstrates that strict priority scheduling protects latency-sensitive Premium workloads at the expense of lower-priority Batch workloads. Aging Priority Scheduling exhibits similar behavior, reducing Premium tenant latency to approximately 76 seconds while increasing Batch tenant latency to approximately 433 seconds. These results indicate that both approaches provide strong QoS guarantees for high-priority tenants under contention.

These results are particularly notable because they are obtained under a workload-characterization strategy that intentionally introduces admission-time estimation errors. Despite operating on coarse workload estimates, Priority and Aging Priority Scheduling continue to preserve strong tenant-level service differentiation, indicating that tenant-priority enforcement remains robust even when workload sizing accuracy is imperfect.

Weighted Scheduling provides a more balanced allocation strategy, allowing Premium workloads to receive preferential treatment while still permitting Standard and Batch tenants to make continuous execution progress. Premium tenant latency is reduced to approximately 158 seconds, while Batch tenant latency remains substantially lower than under strict Priority and Aging scheduling. As a result, Weighted Scheduling offers a practical compromise between fairness and service differentiation.

Similarly, Weighted Scheduling maintains its expected service-allocation behavior despite workload-estimation inaccuracies introduced by the whitespace-based estimator. This observation suggests that tenant-aware scheduling policies are generally less sensitive to workload-characterization fidelity than workload-size-aware schedulers such as SJF.

SJF demonstrates fundamentally different behavior because scheduling decisions are driven primarily by estimated workload size rather than tenant class. Batch workloads experience the lowest average latency (approximately 95 seconds), while Premium workloads experience the highest average latency (approximately 227 seconds). This result indicates that SJF optimizes workload-level efficiency rather than enforcing tenant-level QoS priorities. Because SJF relies directly on admission-time workload characterization, it is also the scheduling policy most sensitive to workload-estimation accuracy. As demonstrated later in the tokenizer-aware evaluation, improving workload-characterization fidelity further improves workload separation and reduces misclassification, reinforcing the effectiveness of workload-size-aware scheduling.

Overall, the results highlight the tradeoff between fairness, latency optimization, and tenant-aware QoS enforcement under the whitespace-based workload-characterization baseline. FIFO maximizes fairness, Priority and Aging maximize Premium tenant protection, Weighted Scheduling balances fairness and service differentiation, and SJF minimizes overall latency by favoring shorter workloads regardless of tenant class. Although workload-characterization fidelity influences absolute latency values, the relative behavior of the scheduling policies remains largely unchanged, indicating that scheduler design exerts a stronger influence on tenant-level QoS outcomes than adaptive bias correction alone.

Figure~\ref{fig:queue_depth_all_schedulers} further illustrates these behaviors through tenant-level queue buildup and drain patterns observed during the experiment. FIFO maintains similar queue evolution across all tenant classes, whereas Priority, Weighted, and Aging Priority exhibit clear service differentiation. SJF displays queue behavior that is primarily determined by workload size rather than tenant category.

\subsection{Queue Waiting Time by Runtime Workload Class}

\begin{table*}[t]
\centering
\caption{Average Queue Waiting Time by Runtime Workload Class Across Workload Characterization and Calibration Modes}
\label{tab:queue_wait_job_type}
\begin{tabular}{lccc|ccc|ccc|cc}
\hline
& \multicolumn{3}{c|}{\texttt{split()}, BIAS=OFF} &
\multicolumn{3}{c|}{\texttt{split()}, BIAS=ON} &
\multicolumn{3}{c|}{Tokenizer, BIAS=ON} &
\multicolumn{2}{c}{Long Wait Change} \\
Scheduler
& Short & Medium & Long
& Short & Medium & Long
& Short & Medium & Long
& OFF$\rightarrow$ON & ON$\rightarrow$Tokenizer \\
\hline
FIFO     & 167.44 & 251.46 & 294.47 & 166.89 & 258.21 & 258.04 & 146.91 & 258.84 & 272.65 & -36.43 & +14.61 \\
Priority & 166.43 & 277.32 & 176.68 & 168.64 & 276.74 &  81.20 & 162.86 & 280.04 & 172.91 & -95.48 & +91.71 \\
Weighted & 160.16 & 265.59 & 227.36 & 168.05 & 265.49 & 164.95 & 167.25 & 262.60 & 244.21 & -62.41 & +79.26 \\
SJF      &   2.89 & 141.06 & 416.28 &   2.87 & 163.18 & 396.59 &   2.93 & 148.79 & 430.32 & -19.69 & +33.73 \\
Aging    & 177.92 & 277.17 & 191.13 & 168.65 & 282.63 &  83.83 & 164.06 & 285.09 & 196.88 & -107.30 & +113.05 \\
\hline
\end{tabular}
\end{table*}

Table~\ref{tab:queue_wait_job_type} summarizes average queue waiting time by runtime workload class across three independent experimental runs using the whitespace-based workload characterization baseline (\texttt{split()}). FIFO scheduling exhibits the expected behavior in which short workloads experience lower queue delay than medium and long workloads while maintaining relatively balanced treatment across workload classes.

SJF produces the strongest workload-size differentiation. Short jobs wait only 2.87 seconds on average, while long jobs wait 396.59 seconds. This confirms that SJF aggressively prioritizes smaller estimated workloads and shifts queueing overhead toward larger jobs. The resulting behavior minimizes latency for short requests but significantly increases waiting time for long-running workloads. Because SJF operates directly on admission-time workload estimates, its effectiveness is highly dependent on workload-characterization fidelity. Under the whitespace-based estimator, workload-size decisions are derived from approximate word-count measurements rather than exact token-space execution cost.

Priority and Aging Priority exhibit a different pattern because tenant priority influences queue ordering more strongly than runtime workload size. Under both policies, long jobs experience substantially lower queue waiting times than medium jobs. This occurs because high-priority tenant requests can be promoted ahead of lower-priority requests regardless of workload size, demonstrating that tenant QoS enforcement dominates workload-size optimization.

Weighted Scheduling provides a compromise between workload-size awareness and tenant-level prioritization. Queue waiting times remain differentiated across workload classes, but the degree of separation is less extreme than under SJF or strict priority-based scheduling.

These results highlight the importance of workload-characterization fidelity. Because SJF and other workload-aware scheduling decisions depend directly on estimated workload size, workload-estimation errors can propagate into queue-ordering decisions and affect scheduler effectiveness. The whitespace-based estimator provides a challenging workload-characterization scenario in which certain requests may be assigned to suboptimal runtime classes. As demonstrated in the tokenizer-aware evaluation, more accurate token-space workload characterization improves workload classification fidelity and reduces admission-time estimation error. Nevertheless, the relative scheduling behavior observed in Table~\ref{tab:queue_wait_job_type} remains consistent, indicating that scheduler design exerts a stronger influence on queue dynamics than workload-estimation refinement alone.

Under Priority and Aging Priority Scheduling, long workloads exhibit lower average queue waiting times than medium workloads. This behavior indicates that tenant-priority assignment dominates workload-size considerations, allowing high-priority long-running requests to bypass lower-priority medium-sized workloads. The result further demonstrates that tenant-aware schedulers optimize service differentiation rather than workload-level efficiency.

\subsection{Impact of Workload-Characterization Fidelity}

\begin{figure*}[t]
\centering

\begin{subfigure}{0.48\textwidth}
\centering
\includegraphics[width=\linewidth]{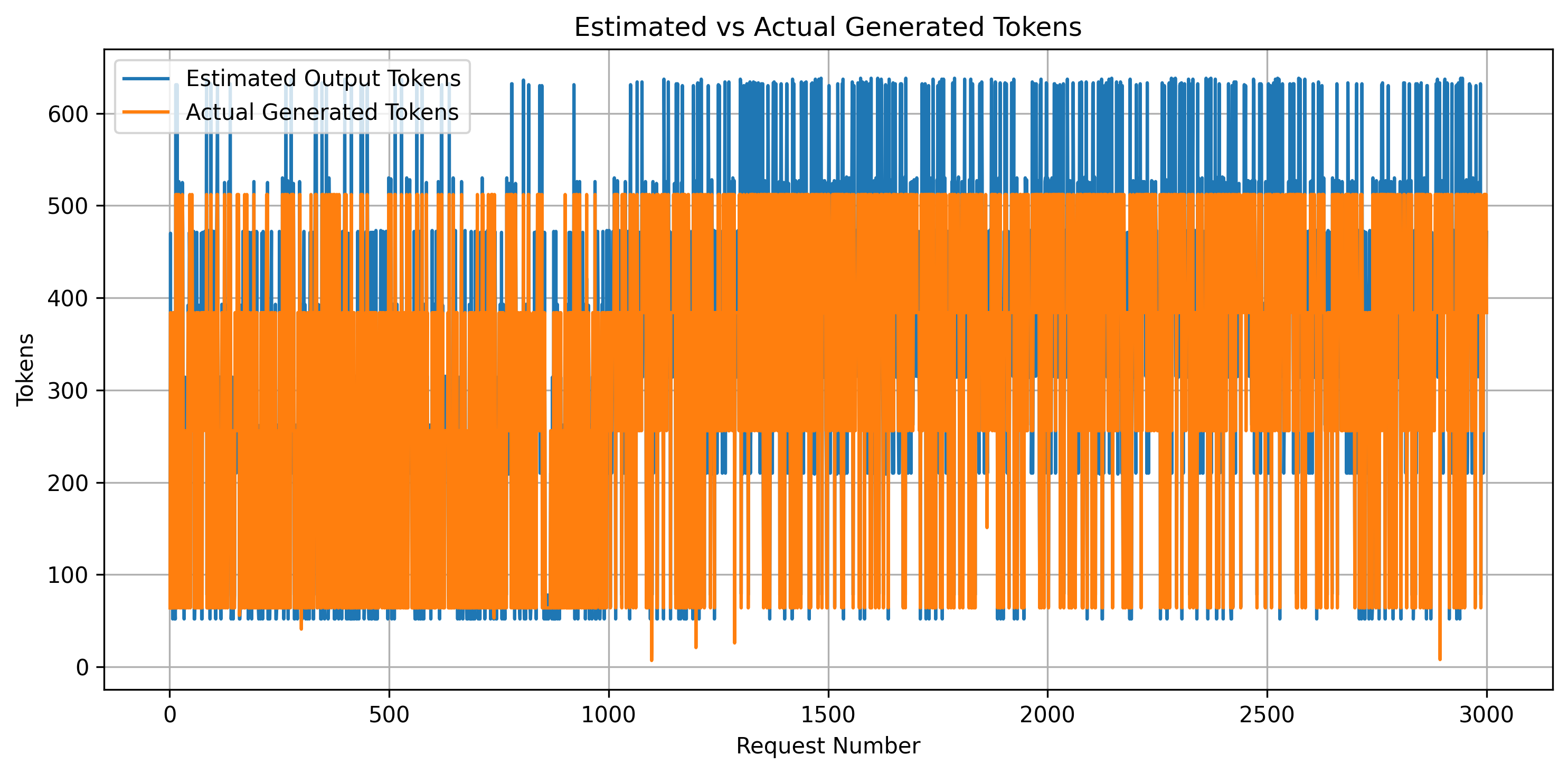}
\caption{FIFO without adaptive runtime token drift compensation}
\label{fig:fifo_no_bias_tokens}
\end{subfigure}
\hfill
\begin{subfigure}{0.48\textwidth}
\centering
\includegraphics[width=\linewidth]{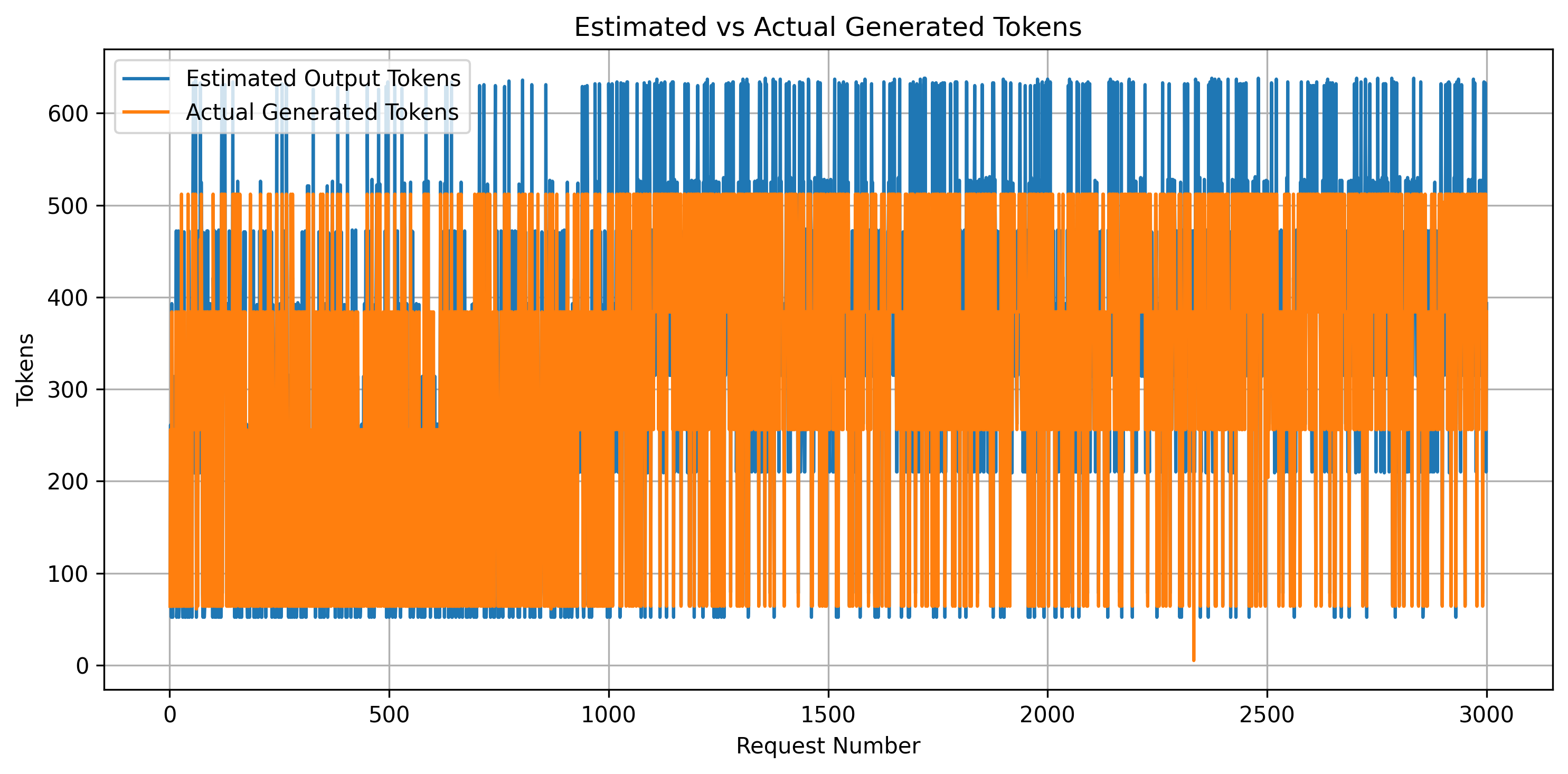}
\caption{FIFO with adaptive runtime token drift compensation}
\label{fig:fifo_with_bias_tokens}
\end{subfigure}

\caption{
Estimated token budgets versus observed output lengths under tokenizer-aware workload characterization with (a) BIAS=OFF and (b) BIAS=ON. Both configurations exhibit nearly identical behavior, indicating that accurate tokenizer-based workload estimation largely eliminates the need for runtime bias correction. Adaptive compensation provides only minor adjustments when admission-time workload characterization is already performed in token space.
}
\label{fig:token_drift_compensation_tokenizer}
\end{figure*}

Figure~\ref{fig:token_drift_compensation} compares admission-time workload estimates and observed output lengths under the whitespace-based workload-characterization baseline (\texttt{split()}) with adaptive calibration disabled (BIAS=OFF) and enabled (BIAS=ON). The figure illustrates how runtime feedback influences workload-estimation behavior when admission-time workload characterization is derived from a coarse word-count proxy rather than exact token-space accounting.

Under the BIAS=OFF configuration, admission-time workload estimates remain fixed and exhibit systematic deviations from observed generation behavior. Because workload budgets are derived from approximate word-count measurements rather than actual tokenizer outputs, the estimator consistently overpredicts runtime execution cost. These workload-characterization errors propagate directly into runtime workload classification decisions and can influence queue ordering behavior, particularly for workload-aware schedulers such as SJF.

When adaptive calibration is enabled (BIAS=ON), workload-specific bias factors are continuously updated using EMA-based runtime feedback. As additional requests complete, estimated workload budgets progressively move closer to observed generation behavior, reducing the mismatch between predicted and actual workload size. This adaptive process improves workload-classification fidelity and partially compensates for the estimation errors introduced by the whitespace-based workload-characterization strategy.

To determine whether adaptive correction remains necessary when workload characterization is performed accurately, we conducted a second evaluation using the model's native tokenizer. Figure~\ref{fig:token_drift_compensation_tokenizer} presents the corresponding tokenizer-aware results under both BIAS=OFF and BIAS=ON configurations.

In contrast to the whitespace-based estimator, tokenizer-aware workload characterization exhibits only minor differences between BIAS=OFF and BIAS=ON. Admission-time workload estimates remain largely unchanged throughout execution, and the adaptive feedback mechanism performs only small corrections. This behavior indicates that most of the estimation error observed under the \texttt{split()} baseline originates from workload-characterization inaccuracies rather than scheduler instability or runtime adaptation limitations.

It is important to note that tokenizer-aware accounting does not eliminate runtime uncertainty entirely. While the tokenizer provides an accurate representation of admission-time token budgets, actual generated output length remains dependent on prompt complexity, response characteristics, stopping conditions, and model behavior. Nevertheless, accurate token-space workload characterization substantially reduces admission-time estimation error and minimizes the need for prolonged calibration.

The results therefore reveal two complementary findings. First, the EMA-based feedback mechanism provides an effective self-healing capability when precise token accounting is unavailable, reducing workload-estimation error and improving workload-classification stability. Second, workload-characterization fidelity exerts a larger influence on estimation accuracy than adaptive correction alone. Once workload size is measured directly in token space, workload-specific bias factors remain close to unity and runtime adaptation provides only marginal additional benefit.

Table~\ref{tab:error_reduction} summarizes workload-estimation improvements observed under the whitespace-based configuration. Across all evaluated scheduling policies, EMA-based calibration reduced workload-estimation error by approximately 39\% in MAE and 40\% in RMSE on average. However, comparison against tokenizer-aware workload characterization demonstrates that accurate admission-time workload estimation provides the greatest overall benefit, reducing workload-classification error before requests enter the scheduling pipeline and improving scheduler decision quality from the outset.

\begin{table}[htbp]
\caption{Average Estimation Error Reduction Across Three Experimental Runs}
\label{tab:error_reduction}
\centering
\small
\begin{tabular}{lcc}
\hline
Scheduler & MAE Reduction & RMSE Reduction \\
\hline
FIFO & 39.51\% & 41.40\% \\
Priority & 39.62\% & 41.36\% \\
Weighted & 38.33\% & 41.10\% \\
SJF & 36.82\% & 37.18\% \\
Aging & 39.74\% & 41.40\% \\
\hline
Average & 38.80\% & 40.49\% \\
\hline
\end{tabular}
\end{table}

The consistent reduction in MAE and RMSE across all evaluated scheduling policies indicates that runtime token drift compensation improves workload estimation independently of the underlying scheduling algorithm.

Adaptive runtime token drift compensation consistently reduced workload estimation error across all schedulers. Averaged across three independent experimental runs, MAE was reduced by approximately 39\% and RMSE by approximately 40\%. Similar improvements were observed across FIFO, Priority, Weighted, SJF, and Aging scheduling policies, demonstrating that the adaptive estimator generalizes across diverse scheduling strategies and workload conditions. The observed reductions remained stable despite the use of an expanded workload corpus containing approximately 1180 unique prompts spanning short question-answering, summarization, technical explanation, and report-generation tasks.

The results demonstrate that workload-aware scheduling policies benefit substantially from accurate workload characterization. Since SJF scheduling decisions depend directly on estimated workload size, tokenizer-aware accounting enables more representative workload classification and improves scheduler effectiveness under multi-tenant load. Runtime calibration further improves robustness when approximate workload-estimation strategies are employed.

\begin{figure}[t]
\centering
\includegraphics[width=\linewidth]{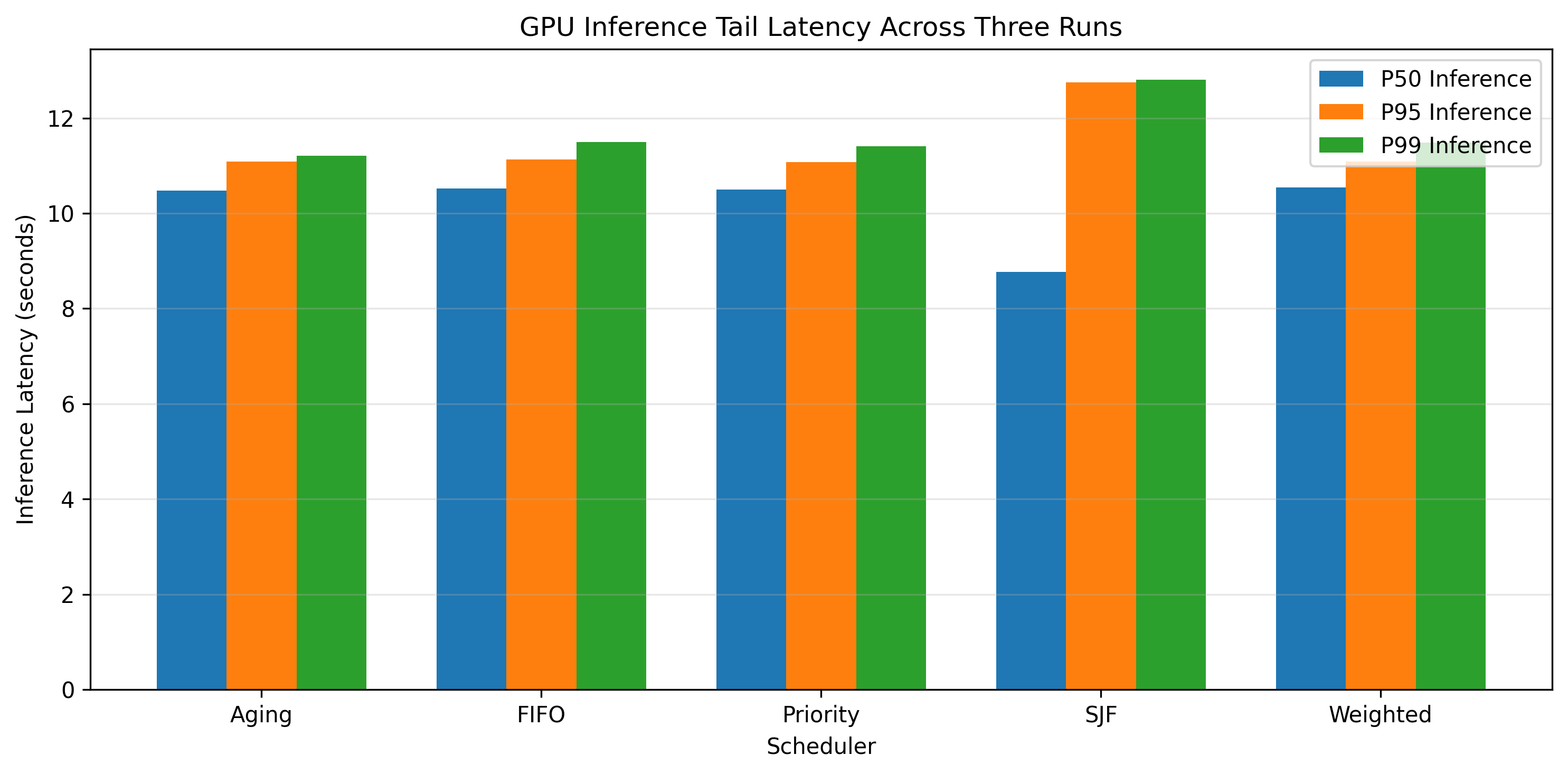}
\caption{GPU latency comparison across scheduling policies.}
\label{fig:gpu_latency_comparison}
\end{figure}

\subsection{Scheduler Comparison}

Figure~\ref{fig:tail_latency_p50_p95_p99} compares end-to-end latency percentiles across all evaluated scheduling policies.

Among all schedulers, SJF achieves the lowest latency across P50, P95, and P99 metrics. Compared with FIFO scheduling, SJF reduces median latency by approximately 42\% while also lowering P95 and P99 latency by more than 15\%. These improvements result from prioritizing smaller workloads and reducing queue occupancy for latency-sensitive requests.

Priority, Weighted, and Aging Priority scheduling provide tenant-aware service differentiation but exhibit latency characteristics similar to FIFO under sustained GPU contention. Although these policies improve tenant-level QoS, they do not significantly reduce overall tail latency.

The results demonstrate that workload-aware scheduling policies benefit substantially from accurate runtime cost estimation. Since SJF scheduling decisions depend directly on estimated workload size, adaptive runtime token drift compensation enables more representative workload classification and improves scheduler effectiveness under multi-tenant load.

\begin{table*}[t]
\centering
\caption{Summary of Key Findings Across Workload Characterization Configurations}
\label{tab:findings_summary}
\begin{tabular}{p{3cm}|p{3.5cm}|p{4cm}|p{5cm}}
\hline
\textbf{Mode} &
\textbf{Workload Estimation Accuracy} &
\textbf{Observed Scheduling Behavior} &
\textbf{Key Finding} \\
\hline

\texttt{split()} + BIAS=OFF &
Coarse workload approximation with systematic estimation error &
Long-job queue waits increase and workload misclassification affects scheduler decisions &
Represents the baseline scenario where admission-time estimation error directly influences scheduling effectiveness \\
\hline

\texttt{split()} + BIAS=ON &
EMA learns workload-specific correction factors and reduces estimation error &
Scheduling behavior converges toward tokenizer-aware configurations after calibration &
Adaptive runtime calibration successfully compensates for persistent workload-estimation errors using lightweight workload characterization \\
\hline

Tokenizer + BIAS=OFF &
Accurate admission-time workload characterization using the native model tokenizer &
Stable latency and QoS behavior without requiring runtime correction &
Accurate tokenization largely eliminates systematic workload-estimation drift and provides reliable workload classification \\
\hline

Tokenizer + BIAS=ON &
Bias values remain near unity throughout execution with only minor fluctuations &
Latency, queue wait, and tenant-level QoS metrics remain nearly identical to Tokenizer + BIAS=OFF &
Adaptive calibration provides limited additional benefit when accurate workload characterization is already available \\
\hline

\end{tabular}
\end{table*}

\subsection{GPU Resource Utilization Analysis}

To determine whether scheduler-dependent performance differences originate from GPU execution efficiency or queue-management behavior, GPU memory consumption and GPU utilization were analyzed across all evaluated scheduling policies.

Results showed that GPU memory consumption remained relatively constant throughout execution, stabilizing near 14.5 GB across FIFO, Priority, Weighted, SJF, and Aging Priority scheduling. This behavior is expected because all experiments used the same model, inference runtime, and continuous batching configuration.

Similarly, GPU utilization remained consistently high across all scheduling policies, typically operating between 85\% and 92\% during sustained execution. Although minor fluctuations were observed due to workload arrival patterns and scheduling decisions, no scheduler produced a substantial change in overall GPU utilization.

These findings indicate that the latency and queue-wait differences observed throughout the study are primarily attributable to workload ordering, queue management, and admission-time scheduling decisions rather than differences in GPU execution efficiency or memory utilization.

\subsection{GPU Execution Latency}

Figure~\ref{fig:gpu_latency_comparison} compares GPU inference latency percentiles across all evaluated scheduling policies. Results are reported using the median (P50), tail (P95), and extreme-tail (P99) latency metrics averaged across three experimental runs.

Unlike end-to-end latency measurements, GPU execution latency remains relatively stable across scheduling policies. FIFO, Priority, Weighted, and Aging Priority scheduling exhibit nearly identical inference latency distributions, with P50 values near 10.5 seconds and P99 values near 11.3 seconds. These observations indicate that the underlying GPU execution cost is largely independent of scheduling policy.

SJF exhibits slightly lower median inference latency while maintaining similar tail latency characteristics. However, the magnitude of improvement is significantly smaller than the reductions observed in end-to-end latency metrics.

The results demonstrate that scheduling policies primarily influence queue management behavior rather than the computational cost of model execution. These observations suggest that the substantial differences in end-to-end latency arise from workload ordering, queue waiting time, and admission-time scheduling decisions rather than changes in GPU processing speed.

This observation highlights the importance of adaptive workload estimation and queue management in multi-tenant inference systems. Improvements achieved by DriftSched are therefore attributable to more effective workload classification and scheduling decisions rather than modifications to the underlying inference runtime.

It is important to note that the results presented in Figure~\ref{fig:gpu_latency_comparison} were obtained using the whitespace-based workload-characterization configuration (\texttt{split()}) with adaptive bias correction enabled. Despite the admission-time workload-estimation inaccuracies introduced by the coarse word-count proxy, GPU execution latency remains nearly identical across all scheduling policies. Additional tokenizer-aware experiments produced comparable GPU inference latency distributions, indicating that workload-characterization fidelity primarily affects admission-time scheduling decisions, queue ordering, and waiting-time behavior rather than the underlying computational cost of model execution itself.

\subsection{Summary of Findings}

Table~\ref{tab:findings_summary} summarizes the primary observations across the four workload-characterization configurations evaluated in this study. The results demonstrate that adaptive runtime calibration is most beneficial when admission-time workload estimation is coarse. Under whitespace-based workload characterization, the EMA-based feedback mechanism successfully compensates for systematic estimation errors and enables scheduling behavior that approaches tokenizer-aware accounting after convergence. In contrast, tokenizer-aware workload characterization largely eliminates admission-time estimation drift, resulting in minimal differences between BIAS=OFF and BIAS=ON configurations. Across all configurations, scheduling-policy selection remained the dominant factor influencing tenant-level QoS differentiation, queue dynamics, and latency performance.

Overall, the experimental results indicate that workload-characterization fidelity influences scheduler behavior primarily when estimation errors are present. However, scheduler selection exerts a substantially larger impact on system performance than workload characterization alone. While tokenizer-aware accounting provides the highest workload-estimation fidelity, adaptive calibration enables lightweight whitespace-based estimators to recover much of the same scheduling behavior after convergence, providing a practical alternative when minimizing admission-path complexity is desirable.

\subsection{Limitations and Future Work}

Several limitations should be considered when interpreting these results. First, under proxy-based workload characterization (\texttt{split()}), the framework requires an initial calibration phase before workload-specific bias factors converge. During this period, admission-time estimation errors may lead to transient increases in queueing delay and tail latency. Although the EMA-based feedback mechanism reduces these errors over time, scheduling effectiveness depends on the availability of sufficient runtime observations.

Second, while tokenizer-aware workload characterization substantially improves estimation accuracy and largely eliminates the need for adaptive correction, it introduces additional preprocessing overhead at the admission layer due to tokenization operations. Under extremely high request arrival rates, this overhead may become a bottleneck and warrants further investigation.

Third, although SJF consistently achieves the strongest latency performance, it may increase starvation risk for long-running requests under sustained contention. While the Aging Priority scheduler mitigates this behavior through priority promotion, additional mechanisms may be required to balance latency optimization and fairness in production deployments.

Future work will investigate adaptive aging strategies, hybrid workload-aware scheduling policies, and scalable admission-layer optimizations to reduce tokenization overhead. Additional evaluation across larger models, multi-GPU environments, and production workload traces would further validate the generality of the proposed approach.

This study was conducted using a single NVIDIA L4 GPU and a single LLM (Qwen1.5-1.8B-Chat). Although the workload corpus contains approximately 1180 prompts across four categories, larger and more diverse workloads may exhibit different token-distribution characteristics. In addition, experiments focus on continuous batching under vLLM and do not evaluate tensor parallelism, multi-GPU deployments, or distributed inference clusters. Future work will investigate larger models, heterogeneous accelerators, and multi-node inference environments.

\subsection{Key Takeaways}

The experimental results lead to four observations.

\begin{enumerate}
\item Scheduler selection has a larger impact on latency than runtime calibration.

\item Workload-characterization fidelity strongly influences admission-time scheduling decisions.

\item EMA-based runtime calibration effectively compensates for systematic errors introduced by lightweight whitespace-based estimation.

\item Tokenizer-aware accounting produces estimates that are already sufficiently accurate, causing calibration factors to remain close to unity and providing little additional benefit from adaptive correction.
\end{enumerate}

\section{Conclusion}

This paper presented DriftSched, a workload-aware QoS scheduling framework for multi-tenant LLM inference serving on shared GPU infrastructure. The framework combines admission-time workload characterization, tenant-aware queue management, multiple scheduling disciplines, and optional runtime calibration through an EMA-based feedback mechanism.

Experimental evaluation on NVIDIA L4 GPUs demonstrated that workload-characterization fidelity plays a critical role in scheduling effectiveness. Comparison between a whitespace-delimited workload proxy and tokenizer-aware accounting showed that inaccurate workload characterization can propagate directly into scheduling decisions, influencing queue dynamics, latency behavior, and workload classification accuracy. Runtime calibration successfully compensates for systematic estimation errors when approximate workload-characterization strategies are employed, while tokenizer-aware accounting substantially reduces the need for runtime correction by aligning admission-time workload estimates with observed execution behavior.

Among the four evaluated configurations, split()+BIAS=OFF produced the largest estimation errors, while split()+BIAS=ON successfully compensated for systematic workload-characterization inaccuracies. Tokenizer-aware accounting achieved substantially lower estimation error without requiring runtime adaptation, and enabling EMA under tokenizer-aware accounting yielded nearly identical results. These findings suggest that workload-characterization fidelity dominates runtime token drift and that adaptive feedback primarily benefits approximate estimators.

Results further showed that scheduling-policy selection exerts a larger influence on end-to-end performance than runtime calibration alone. Among all evaluated schedulers, SJF consistently achieved the strongest latency performance, reducing average queue waiting time and significantly lowering P50, P95, and P99 latency under sustained GPU contention. In contrast, Priority, Weighted, and Aging Priority Scheduling provided stronger tenant-level QoS guarantees at the expense of aggregate latency efficiency. These observations highlight an inherent tradeoff between workload-level latency optimization and tenant-aware service differentiation.

Analysis of GPU execution latency and resource utilization demonstrated that performance differences originated primarily from queue-management behavior and workload ordering rather than changes in GPU execution efficiency. These results indicate that the effectiveness of workload-aware scheduling is driven primarily by improved admission-time decision making rather than modifications to the underlying inference runtime.

This work contributes a workload-aware scheduling architecture, a comparative evaluation of workload-characterization fidelity, a study of five QoS scheduling disciplines, and a reproducible benchmarking framework for multi-tenant GPU inference research. The findings demonstrate that accurate workload characterization is a key enabler of effective QoS-aware scheduling and provide practical guidance for future enterprise-scale AI inference platforms operating under shared GPU contention.

Although tokenizer-aware accounting substantially improves workload characterization, the results consistently show that scheduling policy selection exerts a larger influence on end-to-end latency and QoS behavior than runtime calibration alone.

Future work will investigate adaptive queue reordering, online workload reclassification, multi-GPU scheduling, heterogeneous model serving, and reinforcement-learning-based scheduling approaches capable of dynamically optimizing latency, fairness, and resource utilization under evolving workload conditions.

\end{document}